\documentclass[a4paper,11pt]{article}

\usepackage[top=1in, bottom=1in, left=1in, right=1in]{geometry}

\usepackage{algorithmic}
\usepackage{algorithm}

\usepackage{array}
\usepackage{multirow}
\usepackage{url}
\usepackage{enumitem}
\usepackage{graphicx}
\usepackage[caption=false]{subfig}
\usepackage{rotating}
\usepackage{xcolor}
\graphicspath{{figures/}}

\usepackage{mcode}

\usepackage{hyperref}
\usepackage{datetime}

\newcommand{\sct}{\text{sct}}
\newcommand{\inc}{\text{inc}}
\newcommand{\tot}{\text{tot}}
\newcommand{\diag}{\text{diag}}

\def\sigsize{0.65}
\def\dousize{0.47}
\def\intsize{0.20}

\begin{document}
%
\title{Linearized 3-D Electromagnetic Contrast Source Inversion and Its Applications to Half-space Configurations}

\author{
        Shilong~Sun\thanks{S. Sun, B. J. Kooij, and A. G. Yarovoy are with the Delft University of Technology, 2628 Delft,
        The Netherlands (e-mail: shilongsun@icloud.com; B.J.Kooij@tudelft.nl; A.Yarovoy@tudelft.nl).}
        \and Bert~Jan~Kooij\footnotemark[1]
        \and Alexander~G.~Yarovoy\footnotemark[1]}

\newdate{date}{14}{2}{2017}
\date{\displaydate{date}}

\maketitle

\begin{abstract}

    One of the main computational drawbacks in the application of 3-D iterative inversion techniques is the requirement of solving the field quantities for the updated contrast in every iteration. In this paper, the 3-D electromagnetic inverse scattering problem is put into a discretized finite-difference frequency-domain scheme and linearized into a cascade of two linear functionals. To deal with the nonuniqueness effectively, the joint structure of the contrast sources is exploited using a sum-of-$\ell_1$-norm optimization scheme. A cross-validation technique is used to check whether the optimization process is accurate enough. The total fields are, then, calculated and used to reconstruct the contrast by minimizing a cost functional defined as the sum of the data error and state error. In this procedure, the total fields in the inversion domain are computed only once, while the quality and accuracy of the obtained reconstructions are maintained. The novel method is applied to ground-penetrating radar imaging and through-the-wall imaging, in which the validity and efficiency of the method is demonstrated.

\end{abstract}

\section{Introduction}

    Electromagnetic (EM) inverse scattering is a procedure of recovering the characteristics of the object from the knowledge of the scattered field probed at a limited number of positions \cite{colton2012inverse}. It is of great importance due to the wide spectrum of applications, such as geophysical survey \cite{kuroda2007full,ernst2007full,virieux2009overview,bleistein2013mathematics}, medical diagnosis \cite{gilmore2009microwave,lee2011compressive,rosenthal2013acoustic}, and so forth. Most of the studies on the inverse scattering problems are focused on the frequencies of the resonant region, i.e., the wavelength is comparable to the dimension of the objects. The research related to inverse scattering in the broad sense is still lively today due to the difficulties of dealing with the non-linearity and ill-posedness in the Hadamard sense \cite{hadamard2014lectures}. 

    A variety of inversion methods have been proposed and applied to different applications during the recent decades. Very briefly the methods can be classified into two families --- iterative methods and non-iterative methods. The contrast source inversion (CSI) method is an iterative frequency-domain inversion method to retrieve the value of the contrast (the dielectric parameters of the scattering objects with respect to the background medium) in the testing domain, which was first proposed by P. M. van den Berg et al. \cite{kleinman1992modified,kleinman1993extended,kleinman1994two,van1997contrast}, and was later applied to subsurface object detection in combination with integral equations based on the Electric Field Integral Equation (EFIE) formulation, see Kooij et al. \cite{kooij1999nonlinear}. This idea was further extended to the mixed dielectric and highly conductive objects combined with CSI, see O F{\'e}ron et al. \cite{feron2005microwave}, Chun Yu et al. \cite{yu2005inversion}. To deal with the non-linearity of the inverse problem, Chew proposed the (Distorted) Born iterative methods (BIM and DBIM) \cite{wang1989iterative,chew1990reconstruction,li2004three,gilmore2009comparison}. The key point of this method is to linearize the problem using Born's approximation and consider the reconstructed permittivity as the inhomogeneous background. Although the iterative methods show good performance in achieving the dielectric parameters of the objects, it is extremely time-consuming for the large-scale 3-D inversion scheme with an irregular background due to the fact that each iteration involves the search of finding solutions to the updated scattering problem in the inversion domain. There is another iterative surface-based inversion method, which first parameterizes the shape of the scatterer mathematically with a number of parameters, then sequentially optimizes the parameters by minimizing a cost functional iteratively \cite{lefast}. The drawbacks of this method are obvious. Firstly, it requires a priori information about the position and the quantity of the scatterers. More research on this point can be found in \cite{qing2003electromagnetic,qing2004electromagnetic}. Secondly, it is intractable to deal with the complicated non-convex objects. Linear Sampling Method (LSM) \cite{colton1996simple,colton1997simple} is a non-iterative inversion technique of finding an indicator function for each position in the region of interest (ROI) by first defining a far-field (or near-field \cite{fata2004linear}) mapping operator, and then sequentially solving a linear system of equations. We refer to \cite{catapano2011feasibility} for the application of LSM in ground penetrating radar (GPR). Although LSM has been proven to be effective for highly conductive scatterers, and in some cases, also applicable for dielectric scatterers \cite{arens2003linear}, it is only able to reconstruct the shape of the objects and needs sufficient amount of independent measured data to guarantee the required performance \cite{colton2012inverse}. Besides, it is very time-consuming to compute the dyadic Green functions related to all the voxels in an irregular inhomogeneous background grid \cite{eskandari2014three}.

    Inversion techniques have been investigated mainly in cases where the measured data is obtained from a full aperture setup in order to circumvent the occurrence of local minima in the minimization process of the inversion. However, in many real applications, the ROI can only be illuminated within a very limited range of angles. Among the typical applications are the half-space configurations, for instance, GPR imaging \cite{daniels2005ground} and through-the-wall (TW) imaging \cite{amin2016through}, for which the non-uniqueness is more serious than that of the full aperture cases, because the antennas can only be distributed at a single side of the ROI and only the back-scattered field is available. For solving half-space inverse scattering configurations, the linear focusing methods have been extensively used. For instance, the back-projection method \cite{munson1983tomographic}, time-reversal (TR) technique \cite{fink1993time,fink2000time,micolau2003dort,yavuz2005frequency,liu2007electromagnetic,yavuz2008space,fouda2012imaging,bahrami2014ultrawideband,fouda2014statistical}, its further variant --- TR multiple signal classification (TR MUSIC) \cite{devaney2005time,marengo2006subspace,marengo2007time,ciuonzo2015performance}, and many more. We refer to \cite{catapano2015gpr} for a whole state-of-the-art review. The working principle behind TR imaging is the back propagation of the time-reversed signals observed at the receivers into the imaging region. The process of TR imaging is strictly within the framework of the wave equation, while the back-projection algorithm is a geometrical technique \cite{anastasio2001comments}, which is not based on the wave equation. It is well known that the imaging resolutions of the linear focusing algorithms are bound by the diffraction limit \cite{zhang2013comparison}. In contrast, TR MUSIC became very popular because it is not only algorithmically efficient, but also capable to achieve a resolution that can be much finer than the diffraction limit. As a matter of fact, the method LSM can also be reinterpreted, apart from very peculiar cases, as a ``synthetic focusing'' problem \cite{catapano2007simple}. For 2-D quantitative inverse scattering methods based on LSM and CSI, we refer to \cite{crocco2012linear,di2015inverse,di2015model,rabbani2016hybrid}. A fast 3-D inversion algorithm for solving inverse problems in layered media has been proposed by L. Song and Q. Liu using a so-called diagonal tensor approximation (DTA). This research has been reported in \cite{song2004fast} in which the typical half-space configuration has been discussed. Although this inversion method is efficient, the potential application is limited because of the introduction of an approximate scattering model.  

    In this paper, a linearized 3-D EM contrast source inversion method is proposed and successfully applied to two typical half-space configurations: 1) GPR imaging and 2) TW imaging. Specifically, a finite-difference frequency-domain (FDFD) \cite{W.Shin2013} formulation is used to discretize the forward EM scattering problem, resulting in a highly accurate scattering model, which enables extensive applications to inverse scattering configurations with versatile known backgrounds. With the so-called contrast sources defined as the multiplication of the contrast and the total fields, we formulate the nonlinear inverse scattering problem into a linear model. To deal with the ill-posedness, the contrast sources are estimated by solving a group of $\ell_1$-norm regularized linear problems. A similar idea can be found in the work of G. Oliveri et al. \cite{oliveri2011bayesian}, who have proposed a method in which the contrast sources are obtained separately by a Bayesian compressive sensing method, which is in fact a single measurement vectors (SMV) model because the joint structure of the contrast sources is not considered. In the proposed method, we have exploited the joint structure of the contrast sources by formulating the inverse scattering problem as a linear sum-of-$\ell_1$-norm optimization problem with the multiple measurement vectors (MMV) model \cite{sarvotham2005distributed,chen2006theoretical,van2010theoretical}. The equivalent problem, referred to as the Basis Pursuit denoising (BPDN) problem, is solved instead. We refer to \cite{lee2011compressive} for an application of joint sparsity in the field of medical imaging. As the model is based on a FDFD scheme of the Maxwell equations it enables simple incorporation of complicated background media. In this paper, a 3-D Cartesian coordinate system is used. Therefore, the contrast at each position is described by three coordinates and therefore contains three components. Thus, a group sparse BPDN problem can be obtained, which is solved by a spectral projected gradient for an $\ell_1$ minimization (SPGL1) solver \cite{BergFriedlander:2008,van2011sparse}. Since the noise level is unknown in real applications, a cross-validation (CV) technique \cite{ward2009compressed,zhang2016cross} is used to check whether the optimization process is sufficient. With the estimated contrast sources, the scattered fields can be computed by solving the corresponding forward EM scattering problems. Assuming that the incident fields are known, the total fields can be easily obtained as the summation of the scattered fields and the incident fields. Finally, the contrast is reconstructed by minimizing a cost functional defined as the sum of data error and the state error. The contrast is initialized as the least square solution of the state equations, and the range constraints on the real part and imaginary part of the contrast are considered as a priori information. 

    The proposed method is capable to reconstruct not only the shape but also a coarse estimation of the dielectric parameters of the objects. Since the total fields in the inversion domain are updated only once, this novel method is far more efficient than the traditional iterative inversion methods, e.g., CSI and BIM. We have applied the proposed method to two typical half-space configurations: GPR imaging and TW imaging, and successfully obtained a coarse estimation of the contrast. The forward EM scattering problems are solved by a 3-D FDFD solver --- ``MaxwellFDFD'' and its companion C program ``FD3D'' \cite{maxwellfdfd-webpage}. Moreover, we have also discussed the performance of the proposed linearized inversion method when the dielectric parameters of the background are not exactly known. The remainder of the paper is organized as follows: In Section~\ref{sec.formulation}, the formulation of the inverse scattering problem is introduced. In Section~\ref{sec.EstConSour}, we introduce the reconstruction of the contrast sources, in which the MMV model, the CV-based modified SPGL1 method, and the construction of the scattering matrix are given. The inversion of the contrast is introduced in Section~\ref{sec.IneInvCon}. The inverted results of the numerical experiments of GPR imaging and TW imaging are given in Section~\ref{sec.NumExp} with both exact and inexact background model. Finally, Section~\ref{sec.Conclusion} ends the paper with our conclusions.

\section{Formulation of the Inverse Scattering Problem}\label{sec.formulation}

\begin{figure}[!ht]
        \centering
        \includegraphics[width=0.50\linewidth]{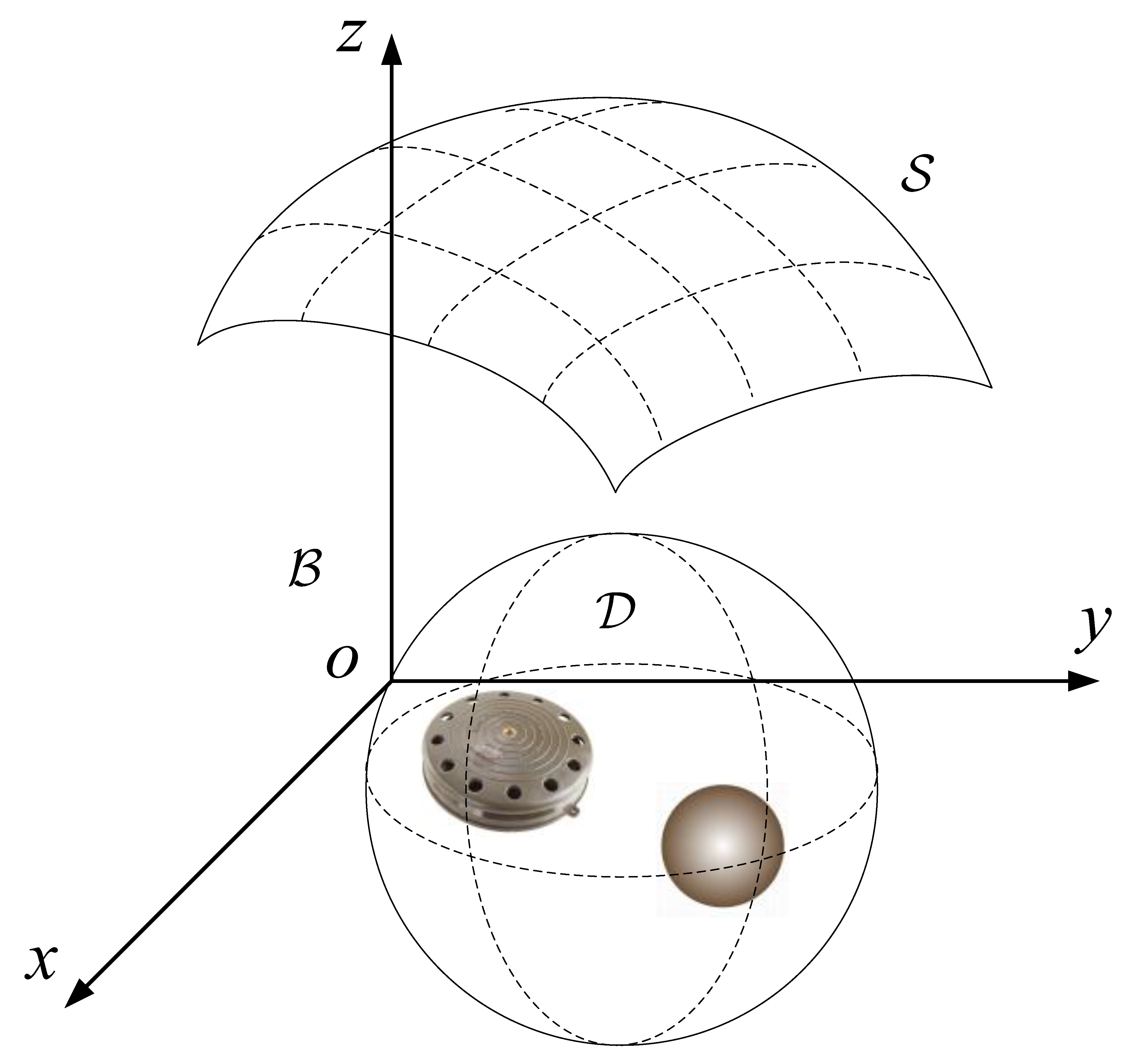}
        \caption{The general geometry of the 3-D inverse scattering problem. Sources and receivers are located on the surface $\mcS$. Objects are located in the inversion region $\mcD$.}
        \label{fig:probSta}
    \end{figure}

    We consider a scattering configuration as depicted in Fig.~\ref{fig:probSta},
    in which sources and receivers are located on the surface $\mcS$, and the objects are located in the background medium $\mcB\subset\mbR^3$. The region $\mcD\subset\mcB$ is the imaging domain which contains the objects. The sources are denoted by the subscript $p$ in which $p\in\{1,2,3...,P\}$, and the receivers are denoted by the subscript $q$ in which $q\in\{1,2,3,...,Q\}$. The electric field corresponding to each source is measured by all the receivers.

    Considering a constant permeability $\mu_0$, we can write the electric field equation in frequency-domain based on the Maxwell's equations as
    \begin{equation}\label{eq.E}
        \nabla\times\mu_0^{-1}\nabla\times\mE_p-\omega^2\vepsilon\mE_p=-\rmi\omega\mJ_p^{\text{src}},\quad p=1,2,\dots,P,
    \end{equation}
    where, $\mE_p$ and $\mJ_p^{\text{src}}$ are the electric field and the electric current source density corresponding to the source with index $p$, respectively; $\omega$ is the angular frequency; $\vepsilon$ is the complex permittivity given by $\vepsilon=\vvarepsilon-\rmi\vsigma/\omega$ with $\vvarepsilon$ and $\vsigma$ representing the permittivity and the conductivity, respectively; $\rmi$ represents the imaginary unit. Note that the time factor is $\exp(\rmi\omega t)$ in this paper. All these quantities are functions of the position vector $\bm{x}$ and the angular frequency $\omega$. 

    Considering the relation of the total fields $\mE_p^{\tot}$, the incident fields $\mE_p^{\inc}$, and the scattered fields $\mE_p^{\sct}$, $\mE_p^{\tot}=\mE_p^{\sct}+\mE_p^{\inc}$, it is easy to obtain the basic equation of the inverse scattering problem, which is denoted by
    \begin{equation}\label{eq.CSI.E}
        \nabla\times\mu_0^{-1}\nabla\times\mE_p^{\sct}-\omega^2\vepsilon_\rmb\mE_p^{\sct} = \omega^2\mJ_p,\quad p=1,2,\dots,P,
    \end{equation} 
    where, $\mJ_p = \vchi\mE_p^{\tot}$ is the contrast source corresponding to the source of the index $p$. Here, $\vchi=\Delta\vvarepsilon-\rmi\Delta\vsigma/\omega$ is the difference of the complex permittivity in the inversion domain with and without the inclusion of the scattering objects and therefore is referred to as the contrast. The aim of this paper is to reconstruct the contrast of the scatterers using the back-scattered field probed on the measurement surface $\mcS$. In the following, we try to fulfill this purpose with measurement data at a single frequency.

\section{Estimation of the Contrast Sources}\label{sec.EstConSour}

\subsection{Formulation with FDFD}

    First, the finite-difference scheme is used in the discretization of the 3-D inverse scattering problem. Since we are only interested in a single frequency inversion, it is advantageous to cast the EM inverse scattering problem into a FDFD scheme. Another advantage is that it is very straightforward to incorporate an inhomogeneous background medium into this FDFD scheme, resulting in applications of our inversion method to different scenarios with inhomogeneous backgrounds.

    Following the vector form of the FDFD scheme in \cite{W.Shin2013}, we recast the vectorial equations Eq.~\eqref{eq.CSI.E} into the following matrix formalism 
    \begin{equation}\label{eq.FD-CSI.eq}
        \mA\ve_p^\sct=\omega^2\vj_p,\quad p=1,2,\dots,P,
    \end{equation}
    where, $\mA\in\mbC^{3N\times 3N}$ is the stiffness matrix in FDFD, which is highly sparse. $\ve_p^\sct\in\mbC^{3N}$ and $\vj_p\in\mbC^{3N}$ are the scattered fields and the contrast sources in the form of a column vector, respectively. Here, $N$ represents the grid number of the discretized 3-D space. Then the solution to Eq.~\eqref{eq.FD-CSI.eq} is obtained by inverting the stiffness matrix $\mA$, which yields
    \begin{equation}\label{eq.fieldsct}
        \ve_p^\sct=\mA^{-1}\omega^2\vj_p.
    \end{equation}
    Considering that the scattered field is measured at a number of positions, we formulate the data equations as follows
    \begin{equation}\label{eq.predata}
        \vf_p= \bm{M}^{\mcS}_p\mA^{-1}\omega^2 \vj_p,\quad p=1,2,\dots,P,
    \end{equation}
    where, $\bm{M}^{\mcS}_p$ is a measurement matrix selecting the values of the scattered fields at the positions of the receivers. Obviously, this is an under-determined linear system of equations. In order to deal with the ill-posedness, the problem of estimating the contrast sources is regularized by the $\ell_1$ norm constraint, which can be formulated as a group of quadratic programming (QP) optimization problems
    \begin{equation}\label{eq.linearmodel}
        \text{(QP$_{\tilde\lambda}$)}\quad
        \underset{\vj_p}{\min}\quad \|\vf_p-\bm{\Phi}_p \vj_p\|_2^2+\tilde\lambda\|\vj_p\|_1, \quad p=1,2,\dots,P,
    \end{equation}
    where, $\bm{\Phi}_p = \bm{M}^\mcS_p \mA^{-1}\omega^2\in\mbC^{M\times 3N}$ is the scattering matrix corresponding to the $p^{\text{th}}$ source. An equivalent problem of (QP$_{\tilde\lambda}$) --- the basis pursuit denoising (BPDN) problem
    \begin{equation}
        \text{(BP$_{\tilde\sigma}$)}\quad
        \underset{\vj_p}{\min}\quad \left\|\vj_p\right\|_1\quad  \text{s. t.} \quad \|\vf_p-\bm{\Phi}_p\vj_p\|_2\leq\tilde\sigma,\quad p=1,2,\dots,P,
    \end{equation}
    is solved instead. As a matter of fact, for appropriate parameter choices of $\tilde\lambda$ and $\tilde\sigma$, the solutions of (QP$_{\tilde\lambda}$) and (BP$_{\tilde\sigma}$) coincide, and the two problems are in some sense equivalent \cite{BergFriedlander:2008}. Since the parameter $\tilde\sigma$ is a measure of the noise level, (BP$_{\tilde\sigma}$) is physically more suitable for the inverse scattering problem. 

\subsection{Group sparse BPND: the MMV model}

    Although the contrast sources $\vj_p$ are excited by the illumination of different incident fields $\ve^\inc_p$, the non-zero values are all located on the surfaces and in the interior of the scatterers. Thus, the contrast sources have the same sparse support, which inspired us to enhance the inversion performance by taking advantage of the joint structure \cite{sarvotham2005distributed}. Hence, the (BP$_{\tilde\sigma}$) problem is further formulated as the sum-of-norms optimization problem \cite{van2011sparse}
    \begin{equation}
        \underset{\mJ}{\min}\quad \sum_{i=1}^{3N}\left\|\mJ_{i,:}^T\right\|_2 \quad \text{s. t.}\quad \left(\sum_{p=1}^{P}\|\vf_p-\bm{\Phi}_p\vj_p\|_2^2\right)^{1/2}\leq\tilde\sigma,
    \end{equation}
    where, $\mJ=\left[\vj_1,\vj_2,\cdots,\vj_P\right]\in\mbC^{3N\times P}$ is a matrix with the contrast sources occupying the columns, and $\mJ_{i,:}^T\in\mbC^{P}$ represents the $i^{\text{th}}$ row of the matrix $\mJ$. Here, $(\cdot)^T$ represents the transpose operator. An interpretation of the sum-of-norms formulation can be simply stated as finding a matrix $\hat{\mJ}$ which has the least value of the sum of the $\ell_1$ norms of the rows while satisfying the inequality.

    Considering the 3-D Cartesian coordinate system, the contrast source at one position consists of three components. If the contrast at this position is not zero, then the three components of the corresponding contrast source are very likely to have non-zero values at the same time. Namely, the contrast sources have a group (group of 3) sparse structure. Therefore, the problem is further formulated as a new sum-of-norms optimization problem
    \begin{equation}
        \underset{\mJ}{\min}\quad \sum_{k=1}^{N}\left\|\left[\mJ_{3k-2,:}\ \mJ_{3k-1,:}\ \mJ_{3k,:}\right]^T\right\|_2 \quad
        \text{s. t.}\quad \left(\sum_{p=1}^{P}\|\vf_p-\bm{\Phi}_p\vj_p\|_2^2\right)^{1/2}\leq\tilde\sigma,
    \end{equation}
    where, $\mJ_{3k-2,:}$, $\mJ_{3k-1,:}$, and $\mJ_{3k,:}$ represent the $x$-, $y$-, and $z$-components of the contrast sources at the position of index $k$, respectively. We solve the sum-of-norms optimization problem with the SPGL1 solver \cite{BergFriedlander:2008,van2011sparse}, of which the basic idea is to find the solution of the BPDN problem by solving a series of Lasso (LS$_\tau$) problems 
    \begin{equation}
        \underset{\vj_p}{\min}\quad \left(\sum_{p=1}^{P}\|\vf_p-\bm{\Phi}_p\vj_p\|_2^2\right)^{1/2}\quad
        \text{s. t.}\quad \sum_{k=1}^{N}\left\|\left[\mJ_{3k-2,:}\ \mJ_{3k-1,:}\ \mJ_{3k,:}\right]^T\right\|_2\leq\tau.
    \end{equation}
    The projection for solving the (LS$_\tau$) problem is implemented by a group projection algorithm (see \cite[Th.~6.3]{van2011sparse}).

\subsection{CV-based modified SPGL1}

    In order to estimate the noise level, i.e., the parameter $\tilde\sigma$, the SPGL1 method is modified based on the CV technique \cite{ward2009compressed,zhang2016cross}. Specifically, we separate the original scattering matrix to a reconstruction matrix $\bm{\Phi}_{p,r}\in\mbC^{M_r\times 3N}$ and a CV matrix $\bm{\Phi}_{p,CV}\in\mbC^{M_{CV}\times 3N}$ with $M = M_r+M_{CV}$. The measurement vector $\vf_p$ is also separated accordingly, to a reconstruction measurement vector $\vf_{p,r}\in\mbC^{M_r}$ and a CV measurement vector $\vf_{p,CV}\in\mbC^{M_{CV}}$. The reconstruction residual and the CV residual are defined as
    \begin{subequations}
        \begin{align}
        \gamma_r &:= \left(\sum_{p=1}^{P}\left\|\vf_{p,r}-\bm{\Phi}_{p,r}\vj_p\right\|_2^2\right)^{1/2},\\
        \gamma_{CV} &:= \left(\sum_{p=1}^{P}\left\|\vf_{p,CV}-\bm{\Phi}_{p,CV}\vj_p\right\|_2^2\right)^{1/2}.
        \end{align}
    \end{subequations}
    In doing so, every iteration can be viewed as two separate parts: 1) reconstructing the contrast sources by SPGL1 and 2) evaluating the outcome by the CV technique, which is utilized to properly terminate the iteration before the recovery starts to over fit the noise. The reconstructed contrast sources are selected as the output on the criterion that its CV residual is the smallest one.

\subsection{Construction of the scattering matrix}

    Note that as the selecting matrix $\bm{M}^{\mcS}_p\in\mbC^{M\times 3N}$ has a small number of rows, the scattering matrix $\bm{\Phi}_p$ can be calculated iteratively by solving $M$ linear systems of equations,
    \begin{equation}\label{eq.linsys}
        \mA^T\vvarphi_{p,m}=(\bm{M}^{\mcS}_{p,m})^T,\quad m=1,2,\dots,M,
    \end{equation}
    where $\bm{M}^{\mcS}_{p,m}$ is the $m^\text{th}$ row of the selecting matrix $\bm{M}^{\mcS}_p$. $\vvarphi_m\in\mbC^{M\times 1}$ constructs the scattering matrix by $\bm\Phi_p = [\vvarphi_{p,1},\vvarphi_{p,2},\dots,\vvarphi_{p,M}]^T$. Since $M$ is much smaller than $N$, the scattering matrix $\bm{\Phi}_p$ is much smaller compared to the LU matrices of the stiffness matrix $\mA$ (if we choose to do LU decomposition for fast calculating the inverse of the matrix $\mA$). This feature makes it possible to compute and store the scattering matrix beforehand, which is of great importance, especially for 3-D inverse scattering problems. It is worth noting that, in the numerical experiments of this paper, the positions of the receivers are fixed for all the measurements, i.e., $\bm{\Phi}_1=\bm{\Phi}_2=\cdots=\bm{\Phi}_P$.

\section{Inversion of the Contrast}\label{sec.IneInvCon}

    Assuming we have reconstructed the contrast sources, the scattered fields can be estimated by Eq.~\eqref{eq.fieldsct}. If we know the incident fields, it is easy to obtain the total fields. In order to solve the contrast, we define the data equations and the state equations as follows
    \begin{subequations}
        \begin{align}
        \label{eq.datsta.a}
        \vf_p & = \bm{\Phi}_p\mD_{\hat e_p^{\tot}}\vchi,\quad p = 1,2,\cdots,P;   \\  
        \label{eq.datsta.b}
        \hat\vj_p & = \mD_{\hat e_p^{\tot}}\vchi,\quad p = 1,2,\cdots,P.
        \end{align}
    \end{subequations} 

    The contrast $\vchi$ can be obtained by iteratively minimizing the cost functional $\mcC(\vchi)$ which is defined as the sum of the data error and the state error
    \begin{equation}
        \underset{\vchi}{\min}\quad \mcC(\vchi):=\frac{\left\|\vf - \bm\Psi\vchi\right\|_2^2}{\left\|\vf\right\|_2^2} + 
        \frac{\left\|\hat\vj-\mD_{\hat e_p^{\tot}}\vchi\right\|_2^2}{\left\|\mD_{\hat e_p^{\inc}}\vchi\right\|_2^2},
    \end{equation} 
    where 
    \begin{equation}
    \vf = 
    \begin{bmatrix}
        \vf_1^T&
        \vf_2^T&
        \hdots&
        \vf_P^T
    \end{bmatrix}^T,
    \end{equation}
    \begin{equation}
    \vj = 
    \begin{bmatrix}
        \vj_1^T&
        \vj_2^T&
        \hdots&
        \vj_P^T
    \end{bmatrix}^T,
    \end{equation}
    \begin{equation}
    \mD_{\hat e_p^{\tot}} = \diag\{\ve_p^{\tot}\},\quad p = 1,2,\cdots,P,
    \end{equation}
    \begin{equation}
    \mD_{\hat e_p^{\inc}} = \diag\{\ve_p^{\inc}\},\quad p = 1,2,\cdots,P,
    \end{equation}
    and 
    \begin{equation}
    \bm\Psi = 
    \begin{bmatrix}
      \bm\Phi_1\mD_{\hat e_1^{\tot}} \\ \bm\Phi_2\mD_{\hat e_2^{\tot}} \\ \vdots \\ \bm\Phi_P\mD_{\hat e_P^{\tot}}
    \end{bmatrix}.
    \end{equation}
    Specifically, $\vchi$ is updated via
    \begin{equation}
    \vchi_n = \vchi_{n-1}+\alpha_n\vnu_{\vchi,n},
    \end{equation}
    where, $\alpha_n$ is a constant and the update direction $\vnu_{\vchi,n}$ is chosen to be the Polak-Ribi{\`e}re conjugate gradient directions given by
    \begin{equation}\label{eq.chinu}
        \begin{split}
            \vnu_{\vchi,0} &= 0\\
            \vnu_{\vchi,n} &= \vg_{\vchi,n}+\frac
                      {
                        \left\langle 
                          \vg_{\vchi,n},\vg_{\vchi,n}-\vg_{\vchi,n-1}
                        \right\rangle
                        _2
                      }
                      {
                          \left\|
                            \vg_{\vchi,n-1}
                          \right\|
                          _2^2
                      }
                      \vnu_{\vchi,n-1}\ n\geq1,
        \end{split}
    \end{equation}
    where, $\vg_{\vchi,n}$ is the gradient of the contrast cost functional $\mcC(\vchi)$ given by
    \begin{equation}\label{eq.gchi}
        \vg_{\vchi,n} = \frac{-2\bm\Psi^H\left(\vf - \bm\Psi\vchi_{n-1}\right)}{\left\|\vf\right\|_2^2} +
        \frac{-2\mD_{\hat e_p^{\tot}}^H\left(\hat\vj-\mD_{\hat e_p^{\tot}}\vchi_{n-1}\right)}{\left\|\mD_{\hat e_p^{\inc}}\vchi_{n-1}\right\|_2^2},
    \end{equation}
    Here, $\left\langle\cdot,\cdot\right\rangle_2$ represents the inner product of two vectors, $(\cdot)^H$ represents the conjugate transpose of a matrix. The step size $\alpha_n$ is determined by minimizing the cost function as follows
    \begin{equation}\label{eq.chicost}
        \mcC_n(\alpha_n)= \frac{\left\|\vf - \bm\Psi(\vchi_{n-1}+\alpha_n\vnu_{\vchi,n})\right\|_2^2}{\left\|\vf\right\|_2^2} + \frac{\left\|\hat\vj-\mD_{\hat e_p^{\tot}}(\vchi_{n-1}+\alpha_n\vnu_{\vchi,n})\right\|_2^2}{\left\|\mD_{\hat e_p^{\inc}}(\vchi_{n-1}+\alpha_n\vnu_{\vchi,n})\right\|_2^2},
    \end{equation}
    which is a problem of finding the minimum of a single-variable function, and can be solved efficiently by using the Brent's method \cite{brent1973algorithms,Forsythe1976computer}.

    The contrast is initialized using the least square solution to the state equations Eq.~\eqref{eq.datsta.b}, i.e., 
    \begin{equation}
        \vchi_0 = \left(\mD_{\hat e_p^{\tot}}\vj\right)\oslash\left(\mD_{\hat e_p^{\tot}}\overline{\hat e_p^{\tot}}\right),
    \end{equation}
    where, $\oslash$ represents the element-wise division.

    By considering the relation, $\vchi=\vepsilon-\vepsilon_b$, where $\vepsilon$ and $\vepsilon_b$ are the complex permittivity of the test domain and the background, and noting the fact that 
    \begin{equation}
        \Re\{\vepsilon\}\succeq 1,\quad \Im\{\vepsilon\}\preceq 0,    
    \end{equation} 
    we can simply obtain
    \begin{equation}\label{eq.rangeconstr}
        \Re\{\vchi\}\succeq 1-\Re\{\vepsilon_b\},\quad \Im\{\vchi\}\preceq -\Im\{\vepsilon_b\}.    
    \end{equation}
    Here, $\Re\{\cdot\}$ and $\Im\{\cdot\}$ represent the real part and imaginary part of a number or a vector; $\succeq$ and $\preceq$ represent the component-wise inequality between the elements of a vector and a constant. Therefore, range constraints are considered in the iterations as a priori information, which is done by setting the real part to $1-\Re\{\vepsilon_b\}$ whenever $\Re\{\vchi_{n}\}<1-\Re\{\vepsilon_b\}$, and setting the imaginary part to $-\Im\{\vepsilon_b\}$ whenever $\Im\{\vchi_{n}\}>-\Im\{\vepsilon_b\}$. In our experiments, the contrast is assumed to be isotropic, i.e., $\chi_{3k-2}=\chi_{3k-1}=\chi_{3k}$. Therefore, we use the mean of the three components as the final estimation of the contrast. 

\section{Numerical Experiments}\label{sec.NumExp}

\subsection{Configuration}

    In this section, the proposed linearized 3-D contrast source inversion method is applied to two typical 3-D half-space inverse problems --- GPR imaging and TW imaging. The forward EM scattering problem is solved by a MATLAB-based 3-D FDFD package 'MaxwellFDFD' and its companion C program ``FD3D''\cite{maxwellfdfd-webpage}. Considering a 3-D Cartesian coordinate system, the $x$-, $y$-, and $z$-normal boundaries of a rectangular  region are covered by perfect matching layers (PML) to simulate the anechoic chamber environment. Non-uniform meshes are used to generate the scattered data, which means the testing domain is discretized with different mesh sizes according to the distribution of the permittivity, viz., coarse meshes for low permittivity and fine meshes for high permittivity. The accuracy of the FDFD scheme is ensured by the following criterion \cite{W.Shin2013}
    \begin{equation}
        \Delta\leq\frac{\lambda_0}{15\sqrt{\varepsilon_r}},
    \end{equation} 
    where, $\lambda_0$ is the wavelength in free space, and $\varepsilon_r$ is the relative permittivity of the testing domain. Non-uniform meshes greatly reduce the computational burden for solving the forward scattering problem. In contrast, uniform meshes are used to invert the scattered data since the distribution of the permittivity is unknown beforehand. In order to guarantee the inverting accuracy, we make sure the following condition is satisfied
    \begin{equation}\label{eq.meshinv}
        \Delta\leq\frac{\lambda_0}{15\sqrt{\max\{\varepsilon_r\}}}.  
    \end{equation} 

    The source used in the numerical experiments consists of an $x$-polarized electric dipole and a $y$-polarized one. A circular polarized wave is generated at 200 MHz by introducing a $\pi/2$ phase shift between the two dipoles. The $x$- and $y$-components of the electric fields are measured at several positions simultaneously. For the two half-space configurations, 6 $\times$ 6 sources are uniformly distributed on the $xoy$ plane $([-3.0,3.0],[-3.0,3.0],z)$ m, and 9 $\times$ 9 receivers are uniformly distributed in the same region. The distance between the receivers both along the $x$- and $y$-axis is $\lambda_0/2=0.75$ m. Here, $\lambda_0$ is the wavelength of the generated wave in free space.

    The measurement data used for inversion consists of the scattered fields obtained by subtracting the incident fields from the probed total fields. Random white noise is added to the measurement data following the similar procedure in \cite{van2003multiplicative}, 
    \begin{equation}\label{eq.noise}
        \vf_{p,noise} = \vf_p+\zeta\times\underset{m}{\max}\{|f_{p,m}|\}(\vn_1+\rmi\vn_2), \quad p = 1,2,\cdots,P,\quad m = 1,2,\cdots,M,
    \end{equation}
    where, $\vn_1$ and $\vn_2$ are two random numbers varying from $-1$ up to 1, $\zeta$ representing the amount of noise, and $\underset{m}{\max}\{|f_{p,m}|\}$ represents the largest value among the amplitudes of the $M$ measurement data, which means the noise is scaled by the largest amplitude of the measurement data. In the following examples, the measurement data is disturbed according to Eq.~\eqref{eq.noise} with $\zeta = 0.05$.

\subsection{GPR imaging: lossy objects}

\begin{figure}[!ht]
        \centering
        \includegraphics[width=\sigsize\linewidth]{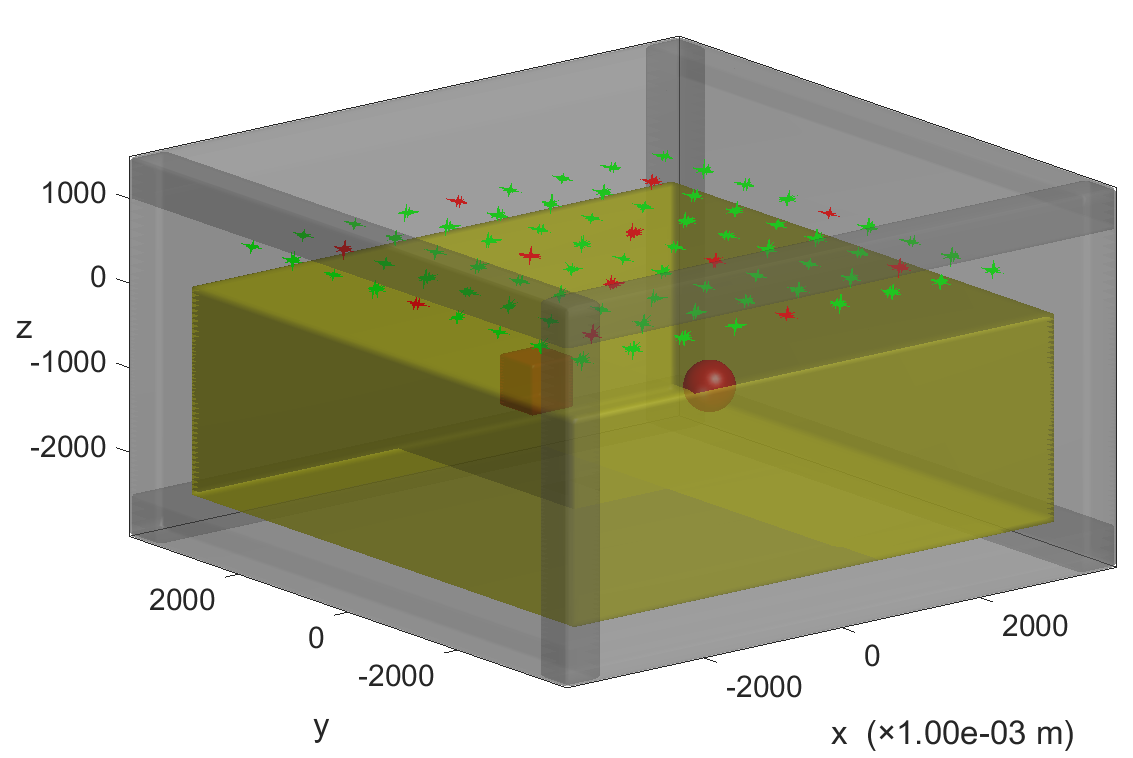}
        \caption{The geometry of the GPR imaging experiment. Soil: $\epsilon_r = 3$, $\sigma = 0.001$ S/m. Sphere: $\epsilon_r = 2$, $\sigma = 0.05$ S/m. Cube: $\epsilon_r = 6$, $\sigma = 0.01$ S/m. 6 $\times$ 6 sources and 9 $\times$ 9 receivers are uniformly distributed on the square plane $([-3, 3], [-3, 3], 0.5)$ m. The 12 red receivers correspond to the CV measurements, and the 69 green ones correspond to the measurements used for reconstructing the contrast sources.}
        \label{fig:confSoil}
    \end{figure}

    In this subsection, we consider the inversion of two lossy objects, a sphere ($\varepsilon_r=2$, $\sigma=0.05$ S/m) of radius 0.3 m and a cube ($\varepsilon_r=6$, $\sigma=0.01$ S/m) of side length 0.6 m, buried in lossy soil ($\varepsilon_r=3$, $\sigma=0.001$ S/m). The testing domain is $([-3.5, 3.5], [-3.5, 3.5], [-2.5, 1.0])$ m. Sources and receivers are uniformly located on the square plane $([-3, 3], [-3, 3], 0.5)$ m, and the half space $-2.5\  \text{m}< z < 0\ \text{m}$ is filled with soil. The sphere is centred at $([0.7,-0.7,-1.0])$ m, and the cube is in the region $([-1.0, -0.4], [0.4, 1.0], [-1.3, -0.7])$ m. Fig.~\ref{fig:confSoil} gives the geometry of this experiment, in which the 9 $\times$ 9 receivers are shown with different colors --- 69 receivers in green color and 12 receivers in red color. The green ones represent the reconstruction measurements and the red ones represent the CV measurements in the optimization process of estimating the contrast sources.

    For creating the scattering matrix $\bm{\Phi}_p$, we discretize the test domain with uniform mesh resolution determined by Eq.~\eqref{eq.meshinv}, and assemble $\bm{\Phi}$ with the vectors $\vvarphi_m$ obtained by solving the $M$ linear system of equations \eqref{eq.linsys}. In order to decrease the computational burden, we constrain the inversion domain in the region $([-2.0, 2.0]$, $[-2.0, 2.0]$, $[-2.5, 0.0])$ m. As a matter of fact, more meshes are used due to the introduction of the PMLs. In our simulations, iterative solvers are used in solving the 3-D forward scattering problems and the scattering matrix. The computation was accelerated by parallel computing programming with 16 cores.  

    \begin{figure}[!ht]
        \centering
        \subfloat[]{\includegraphics[width=0.45\linewidth]{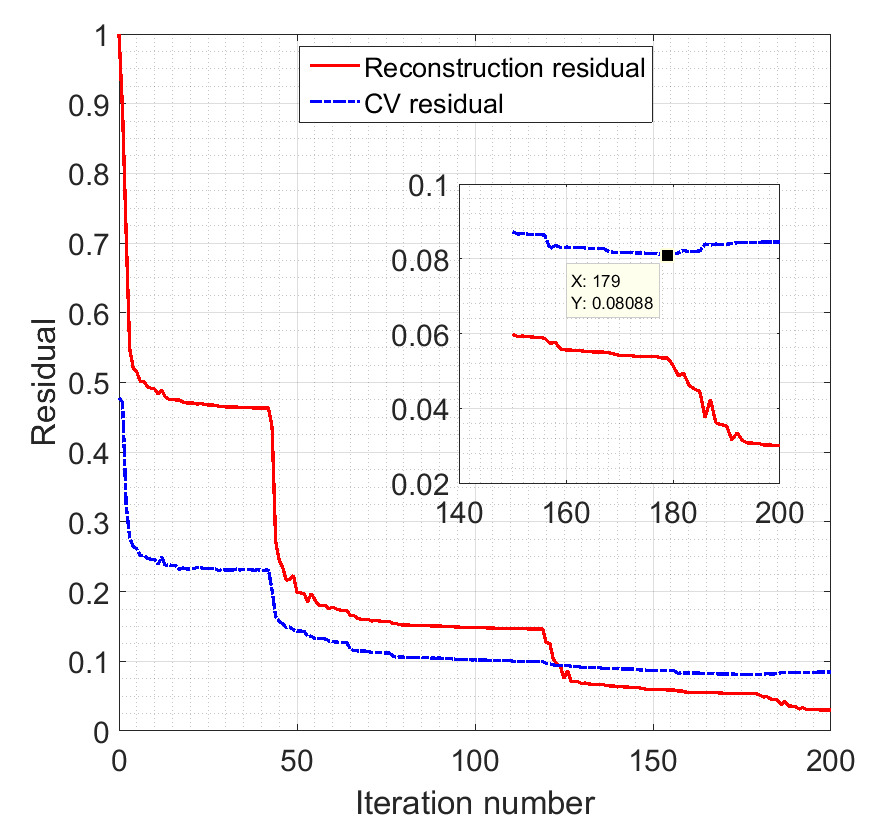}}\quad
        \subfloat[]{\includegraphics[width=0.45\linewidth]{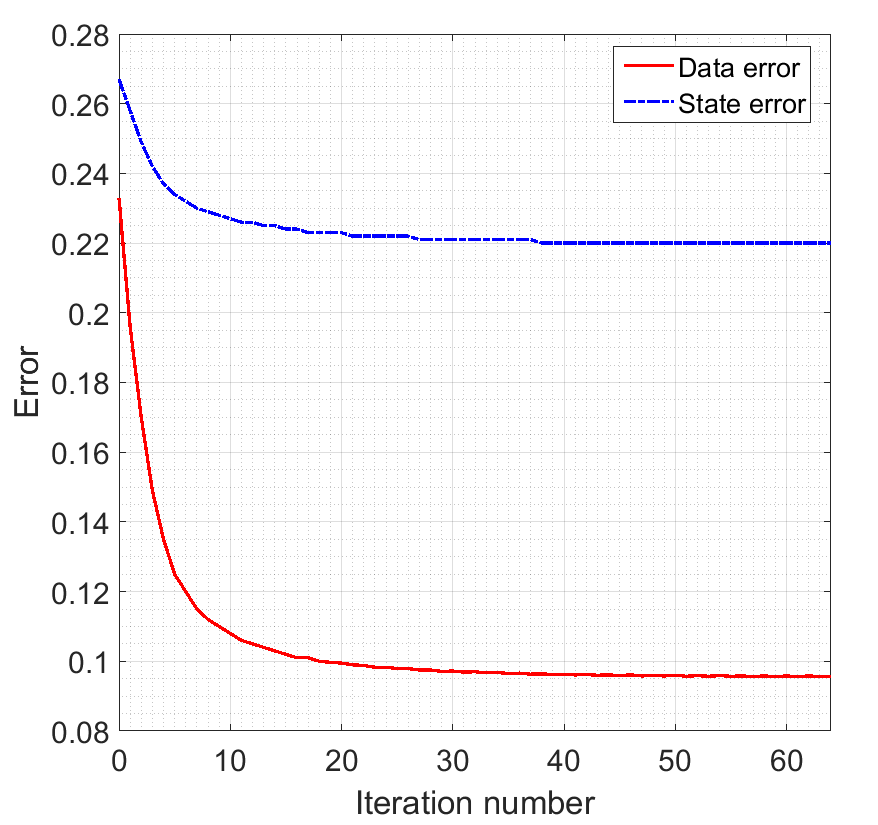}}\\
        \subfloat[]{\includegraphics[width=0.45\linewidth]{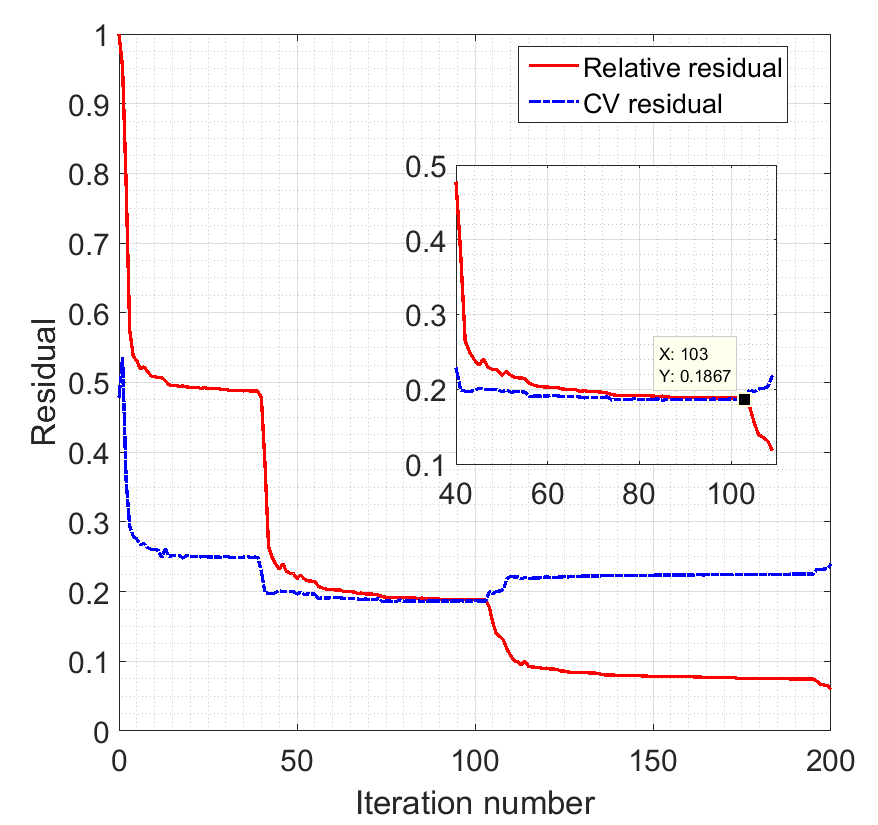}}\quad
        \subfloat[]{\includegraphics[width=0.45\linewidth]{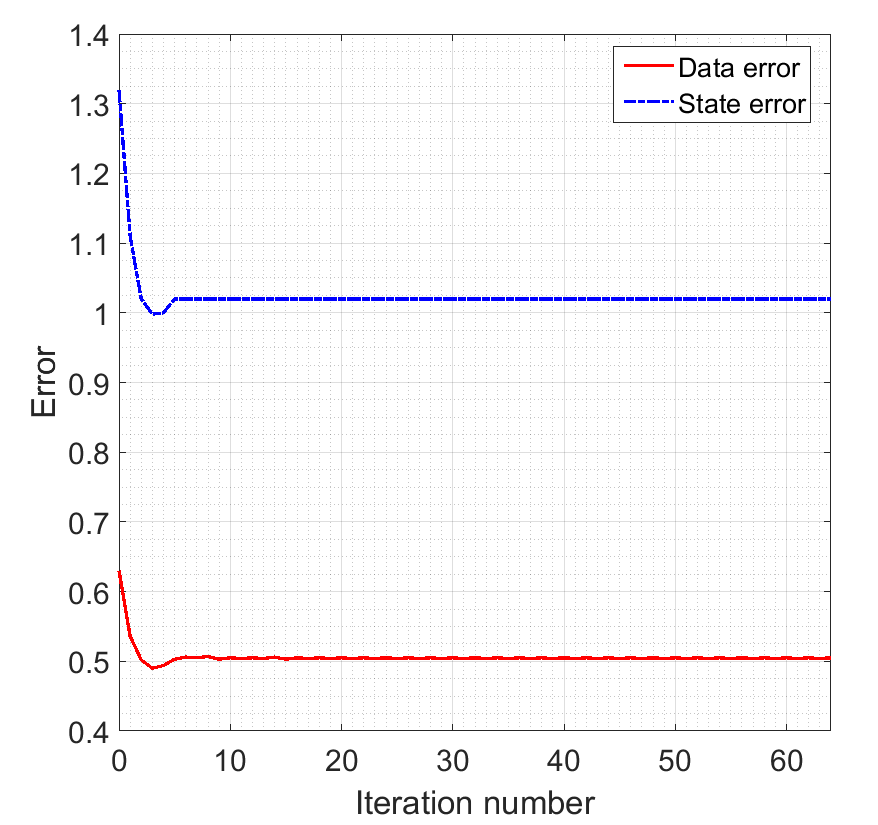}}
        \caption{Residual curves of the GPR imaging experiment. (a) and (c) Reconstruction residual and CV residual curves for estimating the contrast sources using exact background model and inexact background model ($1.25\epsilon_b$), respectively. (b) and (d) Data error and state error curves for reconstructing of the contrast using exact background model and inexact background model ($1.25\epsilon_b$), respectively.}
        \label{fig:CVSoil}
    \end{figure}
    Fig.~\ref{fig:CVSoil}(a) gives the reconstruction residual and the CV residual in the iterative process of estimating the contrast sources, from which we can see that the reconstruction residual curve and the CV residual curve have a stair-like shape. As a matter of fact, the $i^\text{th}$ stair corresponds to the (LS$_\tau$) problem with the parameter $\tau_i$. See \cite{BergFriedlander:2008} for more details about the updating criterion of the parameter $\tau$. From the sub view in Fig.~\ref{fig:CVSoil}(a), we can see that the CV residual starts to increase at the 179$^\text{th}$ iteration, while the reconstruction residual can still be minimized further. This indicates that the iteration process starts to over fit the noise. Therefore, the contrast sources are supposed to be chosen as the approximate solution corresponding to the smallest CV residual. Fig.~\ref{fig:CVSoil}(b) shows the data error curve and the state error curve. From Fig.~\ref{fig:CVSoil}(b) we can see a relatively large state error 22\% has been preserved and cannot be minimized after 40 iterations due to the inexact estimation of the contrast sources, while, in this experiment, the data error has been minimized to a relatively small one 9\%. 

    \begin{figure}[!ht]
        \centering
        \setcounter{subfigure}{0}\subfloat[]
        {\includegraphics[width=\dousize\linewidth] {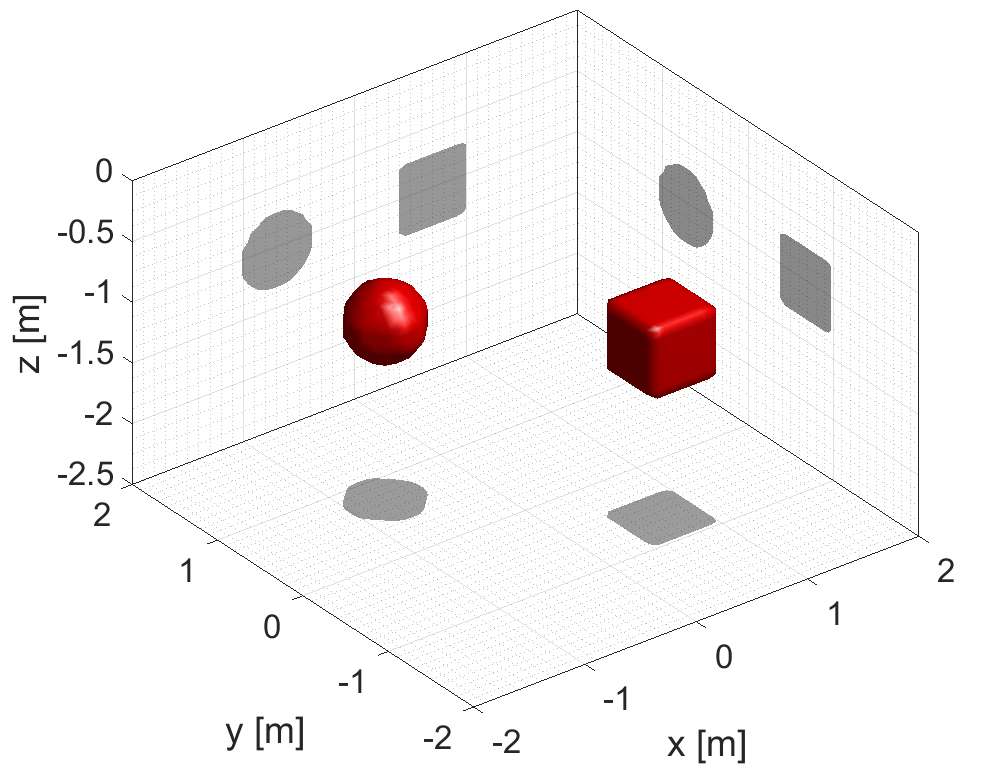}}\hspace*{\intsize cm}
        \setcounter{subfigure}{1}\subfloat[]
        {\includegraphics[width=\dousize\linewidth] {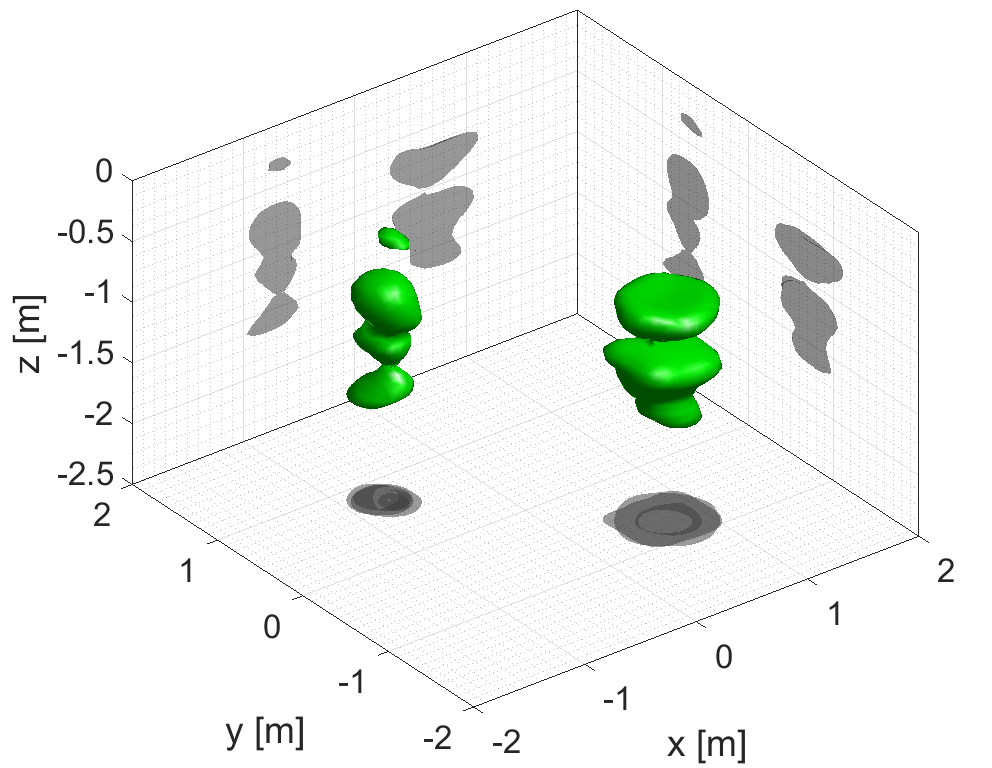}}\\
        \setcounter{subfigure}{2}\subfloat[]
        {\includegraphics[width=\dousize\linewidth] {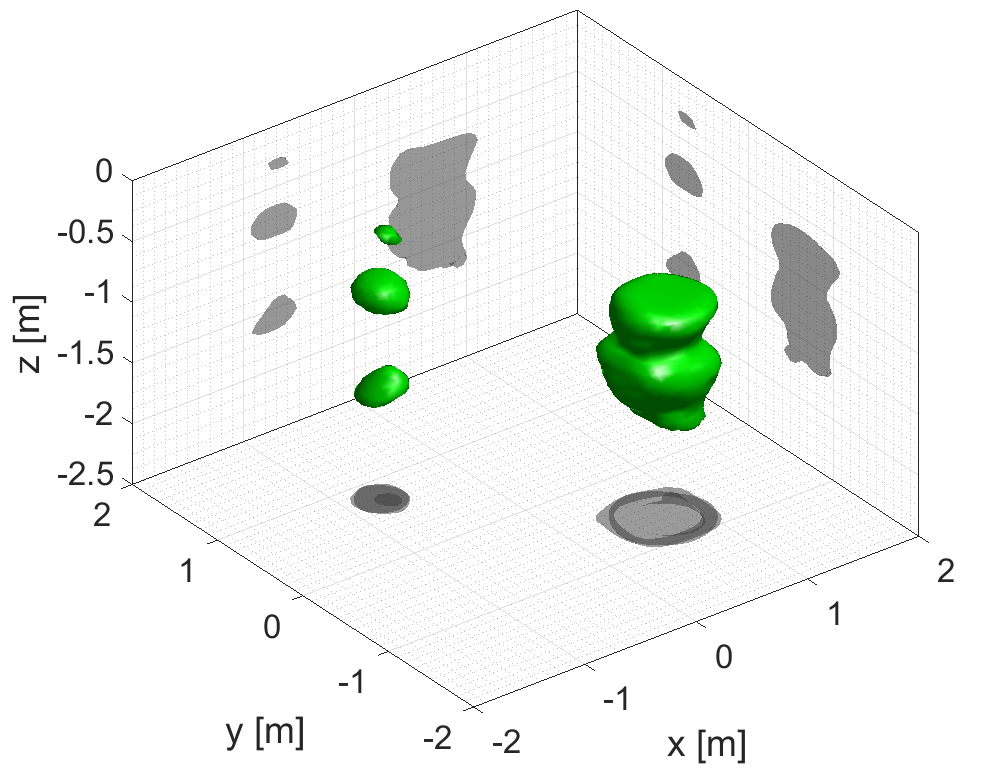}}\hspace*{\intsize cm}
        \setcounter{subfigure}{3}\subfloat[]
        {\includegraphics[width=\dousize\linewidth] {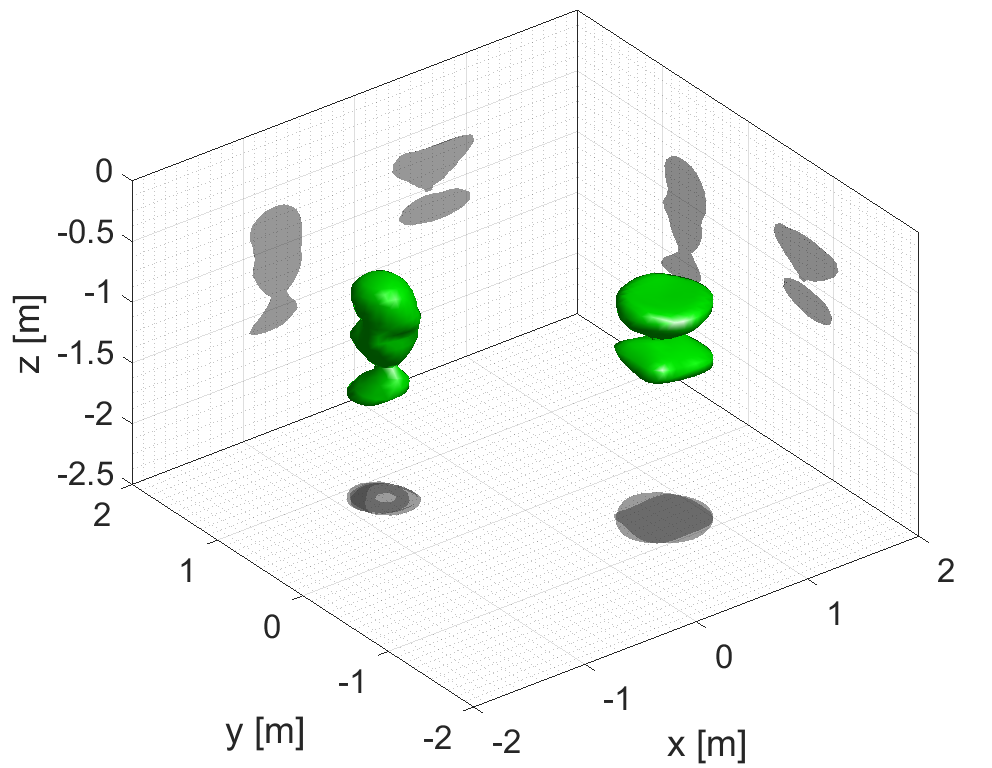}}\\
        \setcounter{subfigure}{4}\subfloat[]
        {\includegraphics[width=\dousize\linewidth] {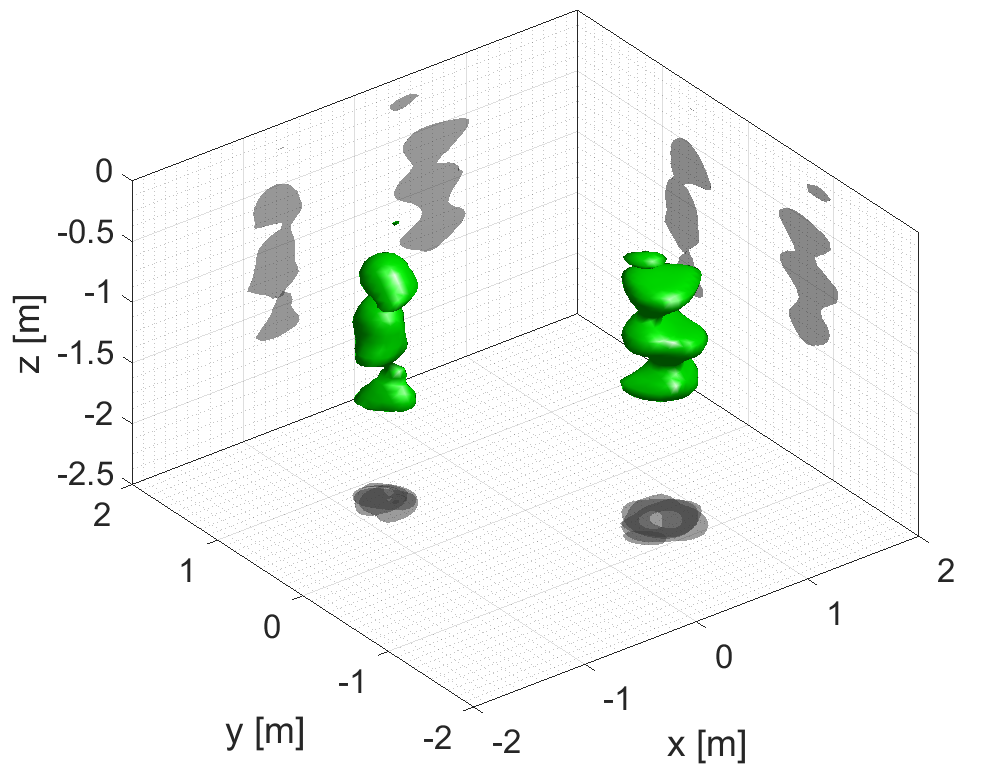}}\hspace*{\intsize cm}
        \setcounter{subfigure}{5}\subfloat[]
        {\includegraphics[width=\dousize\linewidth] {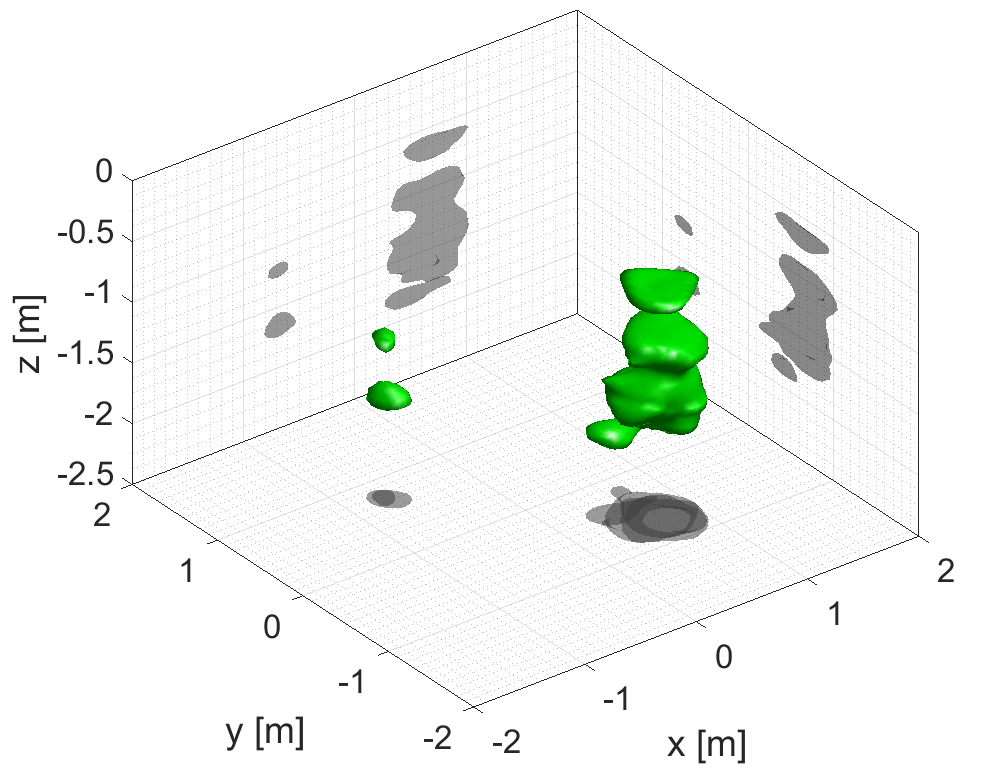}}
        \caption{Three-dimensional shape of the reconstructed results in the GPR imaging experiment at 200 MHz. 5\% random white noise is added. (a) True objects. (b) Reconstructed contrast sources. (c) and (d) Reconstructed contrast permittivity and conductivity using exact background model. (e) and (f) Reconstructed contrast permittivity and conductivity using inexact background model ($1.25\epsilon_b$).}
        \label{fig:invSoilshape}
    \end{figure}
    The shape of the reconstructed results is given in Fig.~\ref{fig:invSoilshape}. Since the contrast is assumed to be isotropic, the shape of the original contrast is defined as 
    \begin{equation}
    I_k = \left|\frac{\chi_{3k-2}+\chi_{3k-1}+\chi_{3k}}{3}\right|,\quad k=1,2,\cdots,N,
    \end{equation}  
    and is shown in Fig.~\ref{fig:invSoilshape}(a). The shape of the contrast sources is defined as 
    \begin{equation}
    I_k = \sum_{p=1}^P\sqrt{|\vj_{p,3k-2}|^2+|\vj_{p,3k-1}|^2+|\vj_{p,3k}|^2},\quad
    k=1,2,\cdots,N,
    \end{equation}  
    and shown in Fig.~\ref{fig:invSoilshape}(b), from which we can see that the contrast sources is elongated along the $z$-axis, indicating good resolution along $x$- and $y$-axis and poor resolution along $z$-axis. This can be explained by the planar distribution of the sources and receivers. Due to the limited aperture in the half-space configurations, the nonuniqueness of the inverse problem gets worse. Thus, it is extremely difficult to do exact reconstruction. Fig.~\ref{fig:invSoilshape}(c) and (d) show the shape of the reconstructed contrast permittivity and the reconstructed contrast conductivity, from which we can see that the location and the basic shape of the buried objects can be well reconstructed. 

    \begin{figure}[!ht]
        \centering
        \setcounter{subfigure}{0}\subfloat[]
        {\includegraphics[width=\dousize\linewidth] {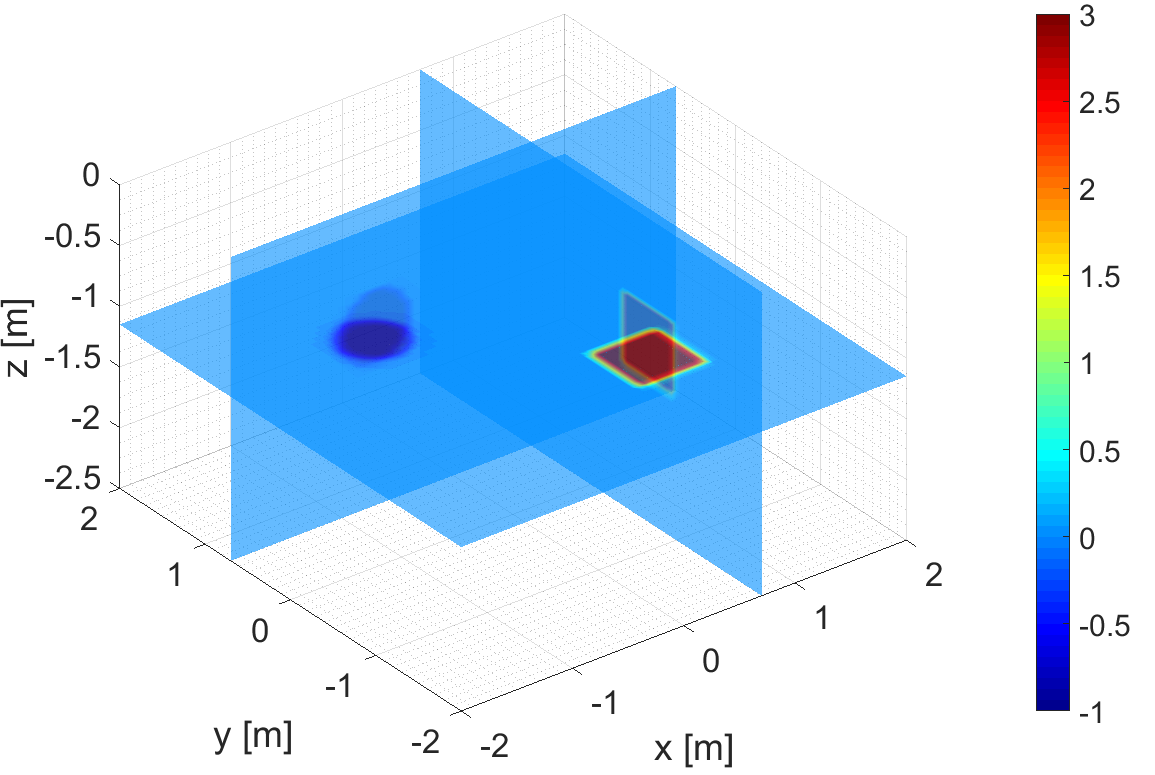}}\hspace*{\intsize cm}
        \setcounter{subfigure}{1}\subfloat[]
        {\includegraphics[width=\dousize\linewidth] {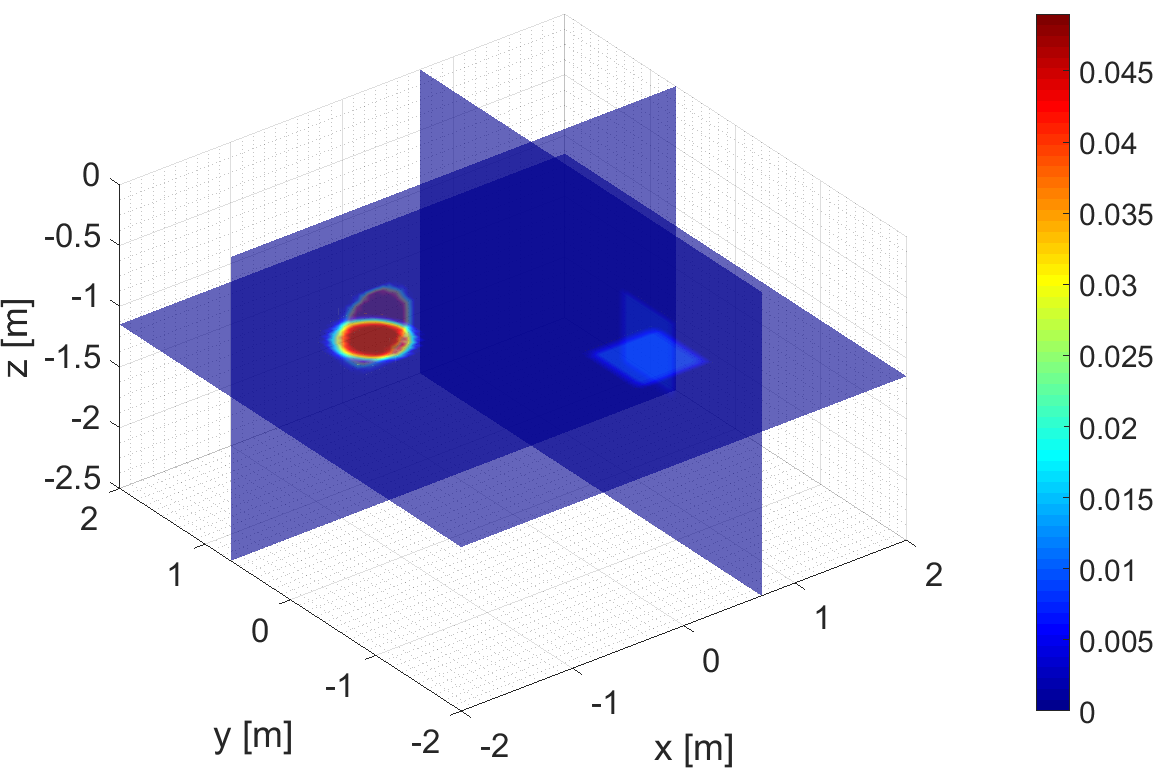}}\\
        \setcounter{subfigure}{2}\subfloat[]
        {\includegraphics[width=\dousize\linewidth] {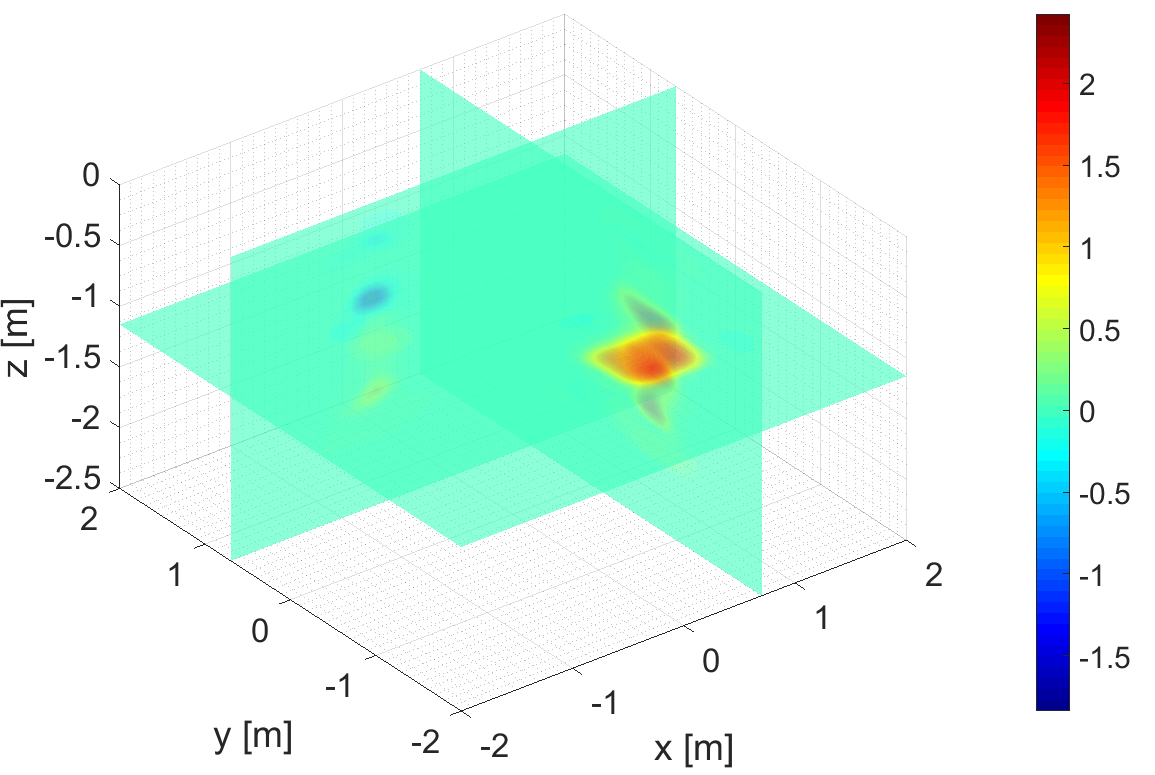}}\hspace*{\intsize cm}
        \setcounter{subfigure}{3}\subfloat[]
        {\includegraphics[width=\dousize\linewidth] {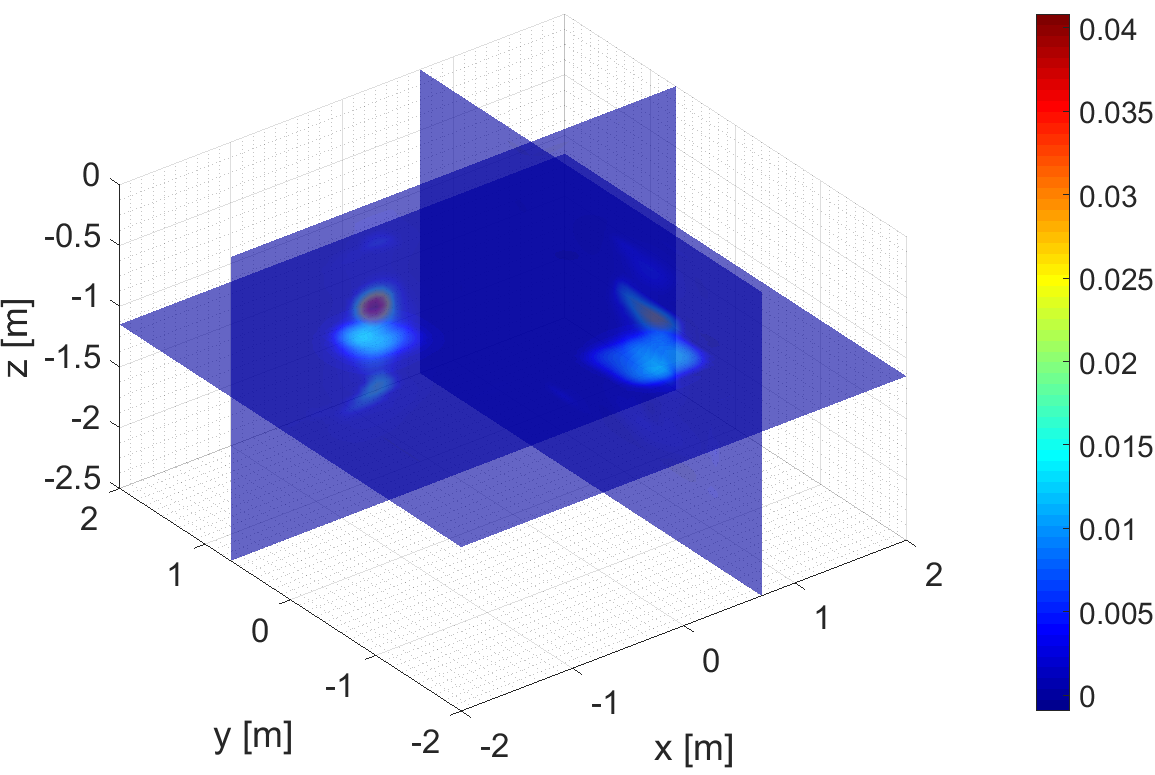}}\\
        \setcounter{subfigure}{4}\subfloat[]
        {\includegraphics[width=\dousize\linewidth] {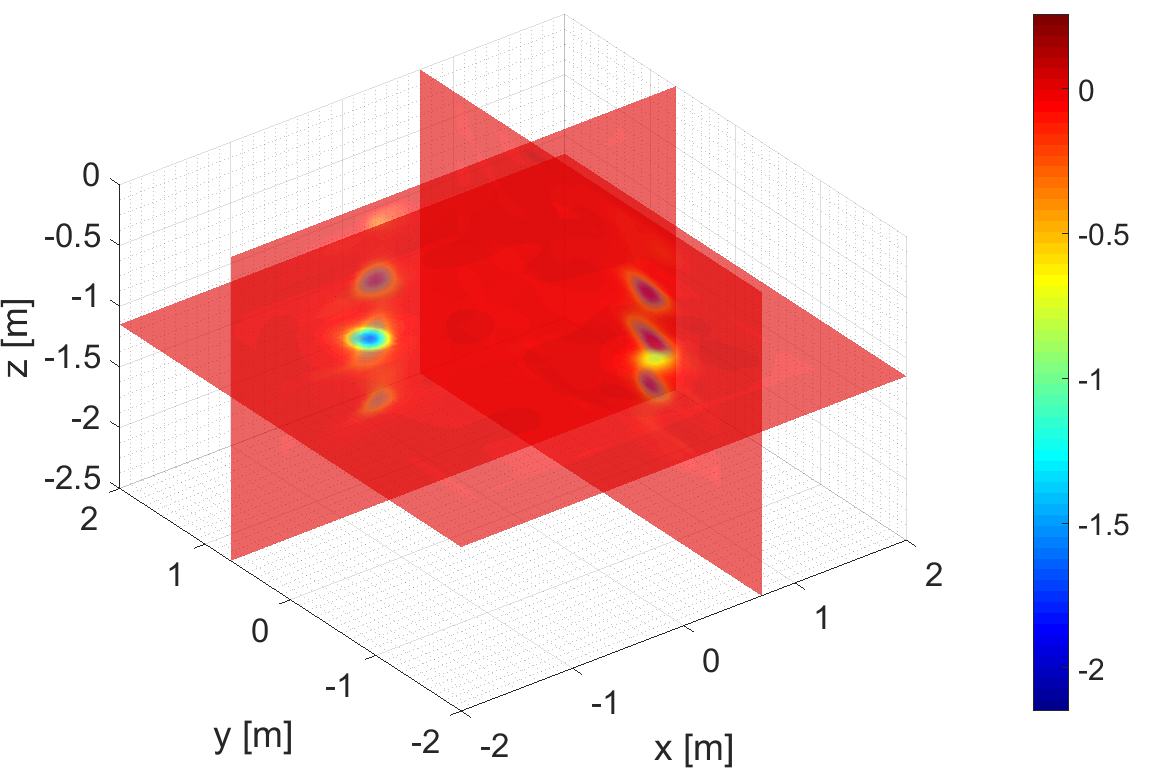}}\hspace*{\intsize cm}
        \setcounter{subfigure}{5}\subfloat[]
        {\includegraphics[width=\dousize\linewidth] {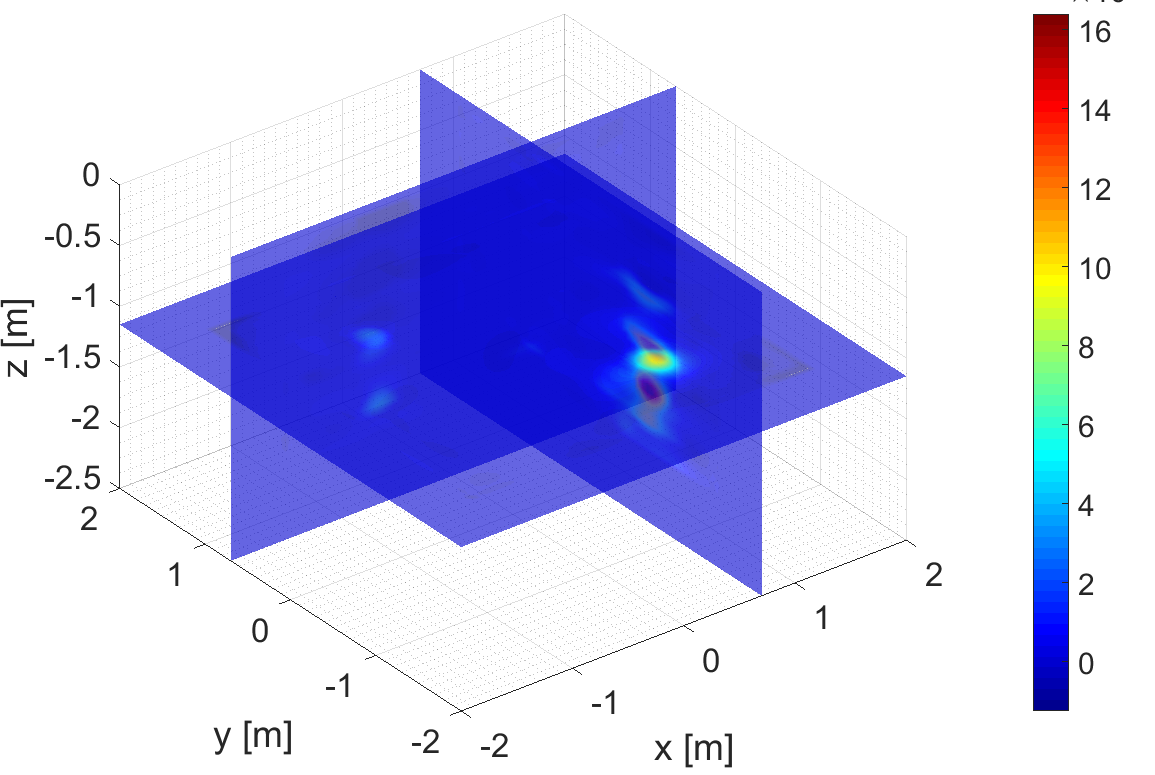}}
        \caption{Cross sections of the reconstructed dielectric parameters in the GPR imaging experiment at 200 MHz. 5\% random white noise is added. The unit of the conductivity is S/m. (a) True contrast permittivity. (b) True contrast conductivity. (c) and (d) Reconstructed contrast permittivity and conductivity using exact background model. (e) and (f) Reconstructed contrast permittivity and conductivity using inexact background model ($1.25\epsilon_b$).}
        \label{fig:invSoil}
    \end{figure}
    To study the estimation accuracy of the dielectric parameters, Fig.~\ref{fig:invSoil} shows the cross sections ($x=0.7$ m, $y=0.7$ m, and $z=-1.15$ m) of the reconstructed contrast permittivity and conductivity, together with those of the real ones. By comparing Fig.~\ref{fig:invSoil}(a) and Fig.~\ref{fig:invSoil}(c), we see that the reconstructed contrast permittivity has negative values down to -1.7 (the real on is -2) in the top region of the sphere, while the one for the cube has positive values up to 2.4 (the real one is 3). Although the estimation of the dielectric parameters is not very accurate, it consists well with the real situation. From Fig.~\ref{fig:invSoil}(b) we see that the sphere has larger conductivity than the cube, this is well presented in the reconstructed results shown in Fig.~\ref{fig:invSoil}(d), and the maximum value of the estimated contrast conductivity is 0.04 S/m which is very close to the real value 0.05 S/m. 

    In order to study the influence of the background mismatch to the inversion performance of the proposed method, we process the same measurement data with an inexact background model. Specifically, we assume the geometry of the ground is exactly known, but the dielectric parameters of the soil are estimated higher than the exact value by 25\%. It is worth noting that the incident fields, as well as the scattering matrix $\bm{\Phi}_p$, have to be recalculated according to the inexact background model. In addition, the contrast must be restricted according to the newly estimated dielectric parameters of the soil. Fig.~\ref{fig:invSoilshape}(e) and (f) gives the shape of the inverted results obtained by processing the same measurement data with the inexact dielectric parameters of the soil ($1.25\times\vepsilon_b$). From Fig.~\ref{fig:invSoilshape}(e) and (f) we can see that more artefacts have been reconstructed due to the mismatch of the background model. The resolution in the $x$- and $y$-dimension is still acceptable, but the resolution in the $z$-dimension gets worse. Fig.~\ref{fig:invSoil}(e) and (f) gives the cross sections ($x=0.7$ m, $y=0.7$ m, and $z=-1.15$ m) of the reconstructed contrast permittivity and conductivity, from which we can see that the mismatch seriously degrades the reconstruction accuracy about the dielectric parameters of the objects.

\subsection{TW imaging}

    \begin{figure}[!ht]
        \centering
        \includegraphics[width=\sigsize\linewidth]{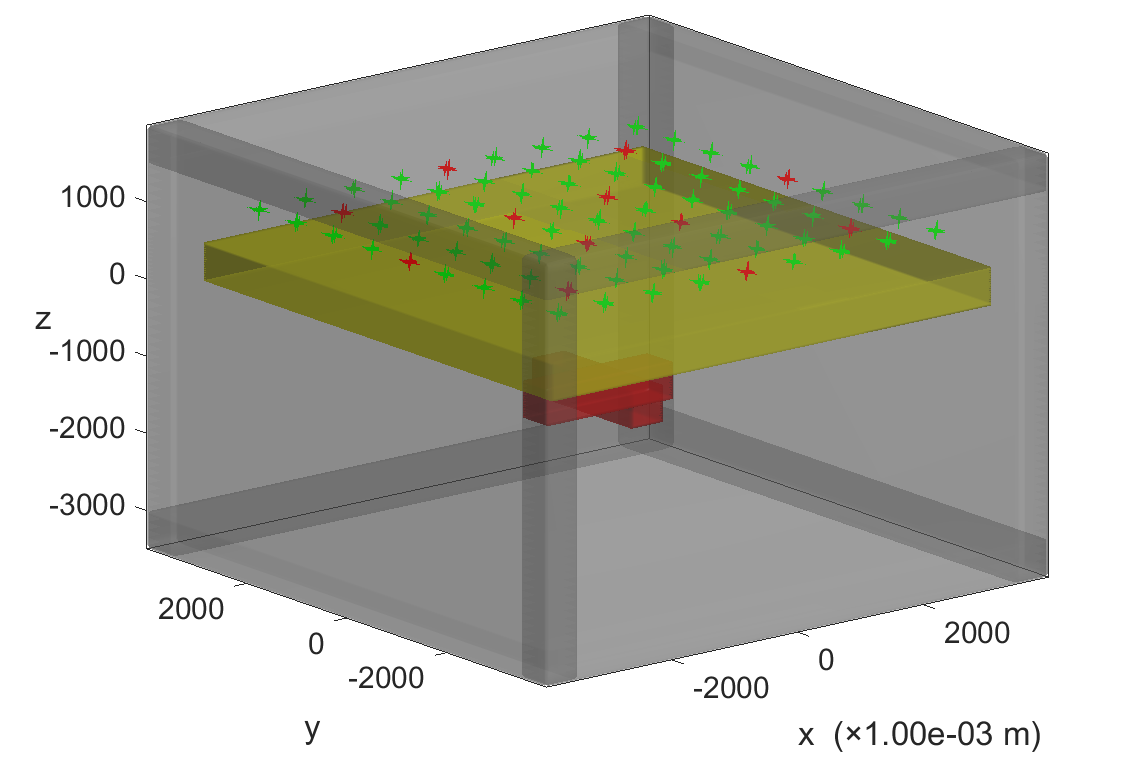}
        \caption{The geometry of the TW imaging experiment. Wall: $\epsilon_r = 4$, $\sigma = 0.01$ S/m. Object: lossy material $\epsilon_r = 2$, $\sigma = 0.001$ S/m and highly conductive material $\epsilon_r = 1$, $\sigma = 1$ S/m. The wall is in the region $([-3.5, 3.5], [-3.5, 3.5], [0, 0.5])$. 6 $\times$ 6 sources and 9 $\times$ 9 receivers are uniformly distributed on the square plane $([-3, 3]; [-3, 3]; 1.0)$ m. The 12 red dots represent the CV measurements, and the 69 green dots are the reconstruction measurements.}
        \label{fig:confwall100sm}
    \end{figure}
    In this subsection, we consider the inversion of a cross object placed behind a wall ($\varepsilon_r=4$, $\sigma=0.01$ S/m). The testing domain is $([-3.5, 3.5], [-3.5, 3.5], [-3.0, 1.5])$ m. 6 $\times$ 6 sources and 9 $\times$ 9 receivers are uniformly located on the square plane $([-3, 3], [-3, 3], 1.0)$ m, and the wall is in the region $([-3.5, 3.5], [-3.5, 3.5], [0, 0.5])$. The cross object is combined with two rectangular blocks in the region $([-1.0, 1.0], [-0.25, 0.25], [-1.5, -1.0])$ m and $([-0.25, 0.25], [-1.0, 1.0], [-1.5, -1.0])$ m. The geometry of the TW imaging experiment is shown in Fig.~\ref{fig:confwall100sm}. The receivers are shown with different colors, of which the green ones represent the reconstruction measurements and the red ones represent the CV measurements. The inversion domain is constrained in the region $([-2.0, 2.0], [-2.0, 2.0], [-2.5, 0.0])$ m. 

    In this TW imaging experiment, we investigate the inversion performance of the proposed method not only to the lossy object, but also to the highly conductive object. For the latter, the morphological information is of more interest. The exact and inexact wall models are all considered for solving the scattering matrix $\bm\Phi_p$ and modelling the incident fields.

\subsubsection{Lossy object}

    \begin{figure}[!ht]
        \centering
        \subfloat[]
        {\includegraphics[width=0.45\linewidth]{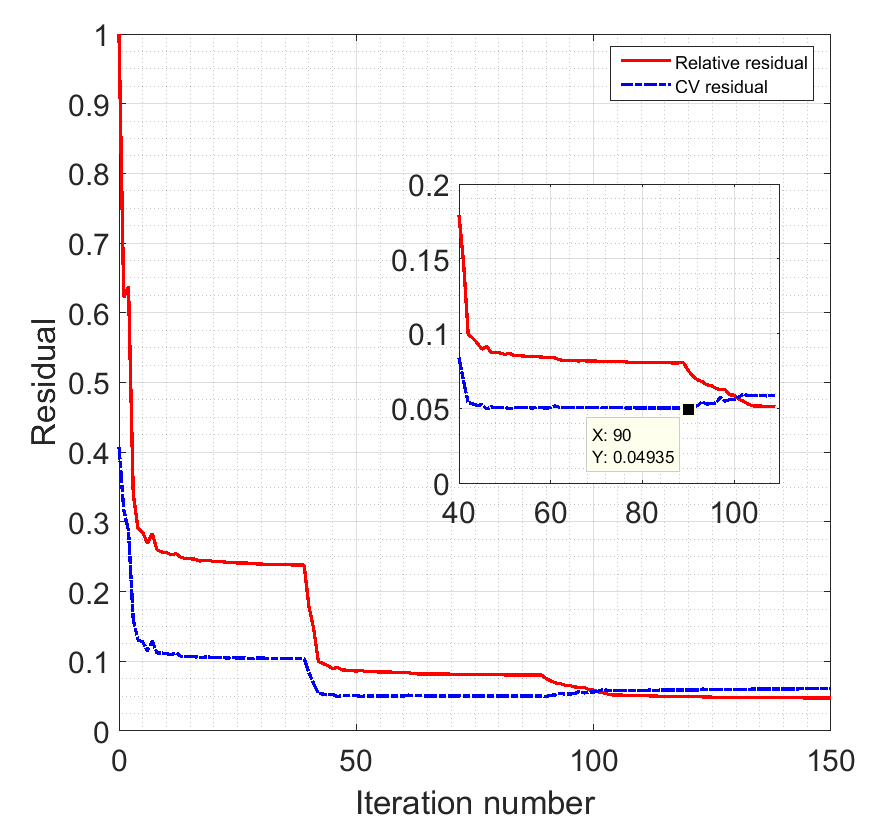}}\quad
        \subfloat[]
        {\includegraphics[width=0.45\linewidth] {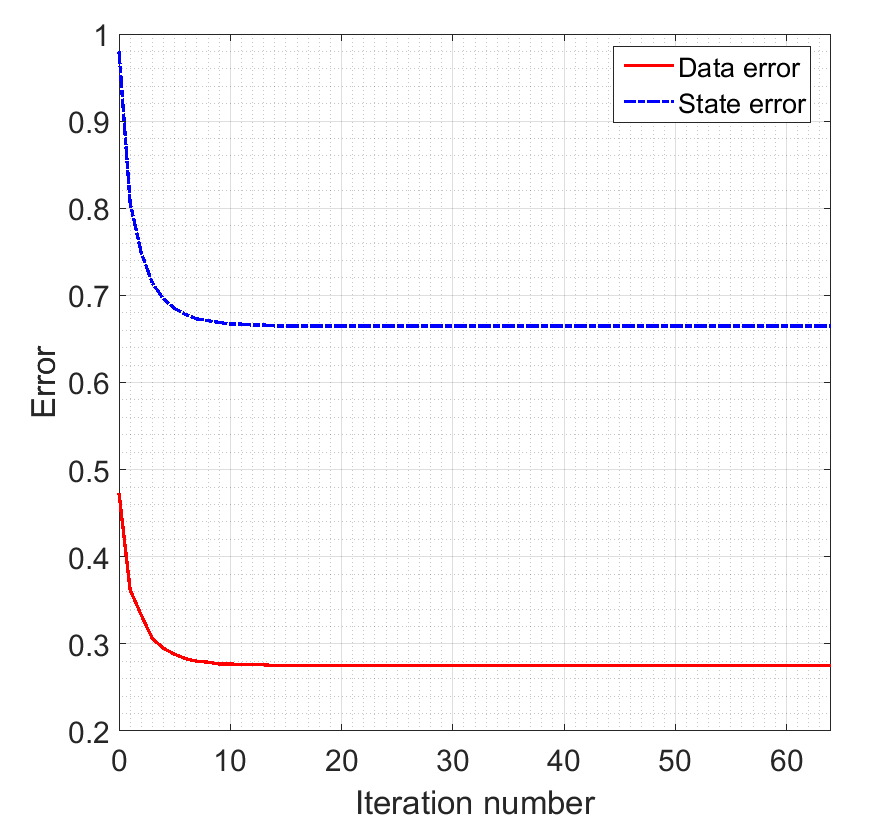}}\\
        \subfloat[]
        {\includegraphics[width=0.45\linewidth]{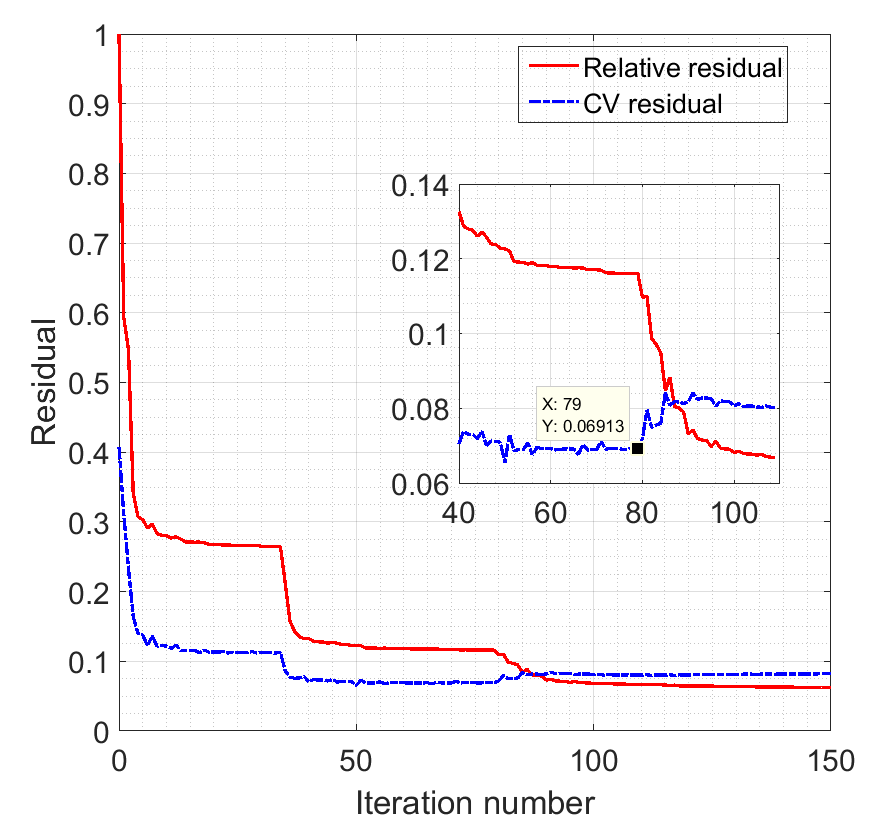}}\quad
        \subfloat[]
        {\includegraphics[width=0.45\linewidth] {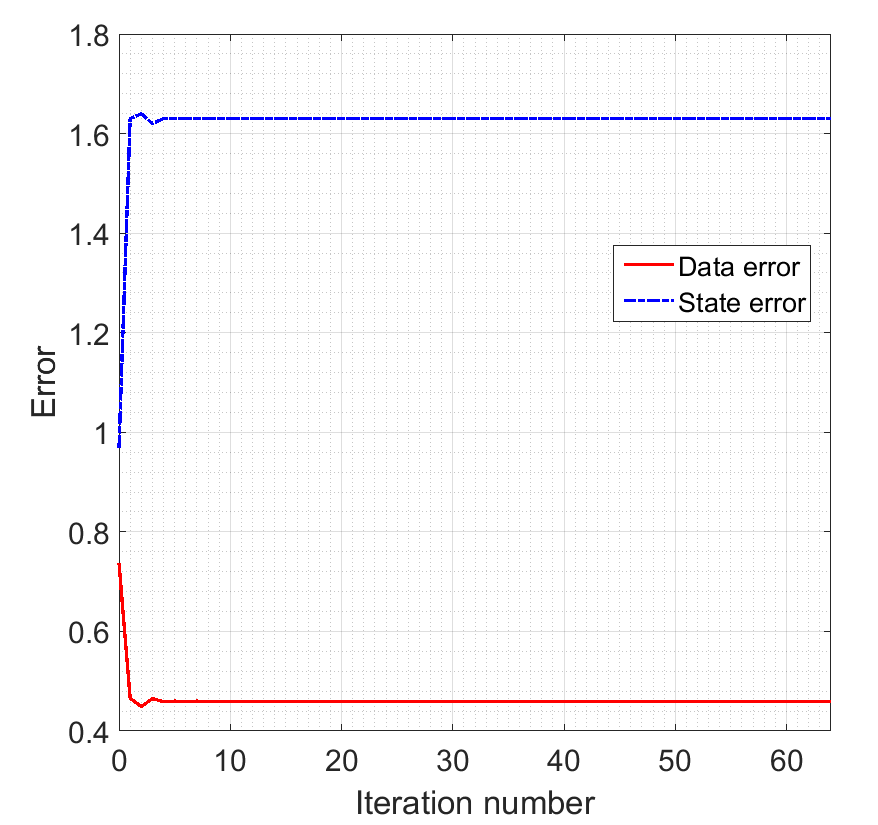}}
        \caption{Residual curves of the TW imaging experiment with lossy object. (a) and (c) Reconstruction residual and CV residual curves for estimating the contrast sources using exact background model and inexact background model ($0.75\epsilon_b$), respectively. (b) and (d) Data error and state error curves for reconstructing the contrast using exact background model and inexact background model ($0.75\epsilon_b$), respectively.}
        \label{fig:CVwalldie}
    \end{figure}
    \begin{figure}[!ht]
        \centering
        \setcounter{subfigure}{0}\subfloat[]
        {\includegraphics[width=\dousize\linewidth] {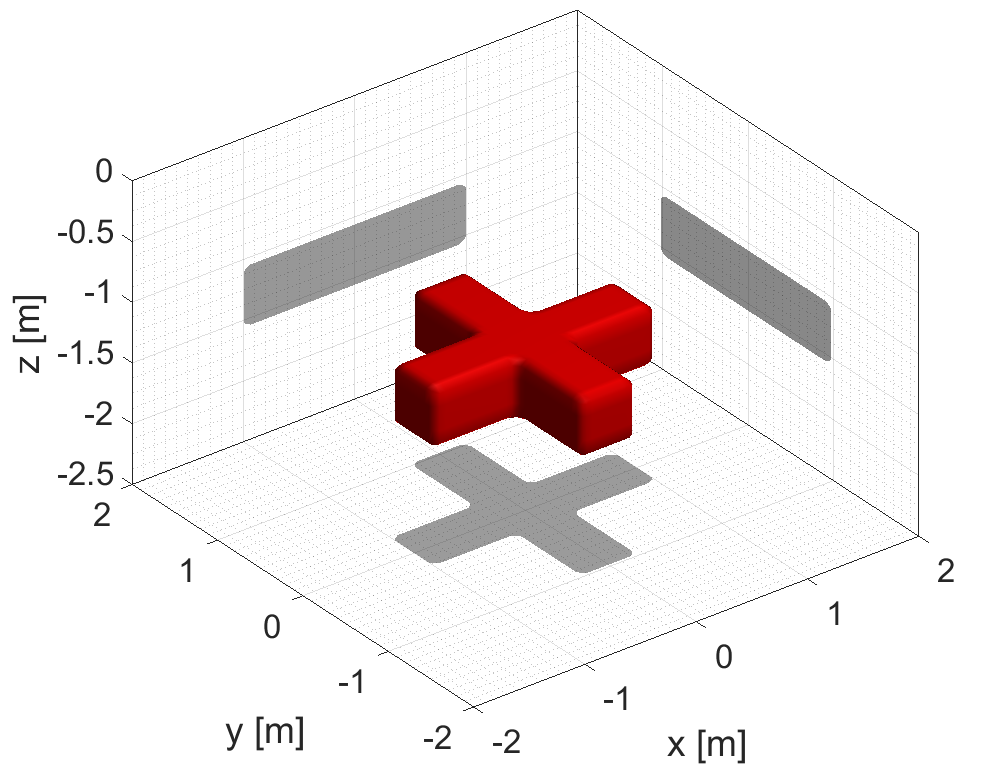}}\hspace*{\intsize cm}
        \setcounter{subfigure}{1}\subfloat[]
        {\includegraphics[width=\dousize\linewidth] {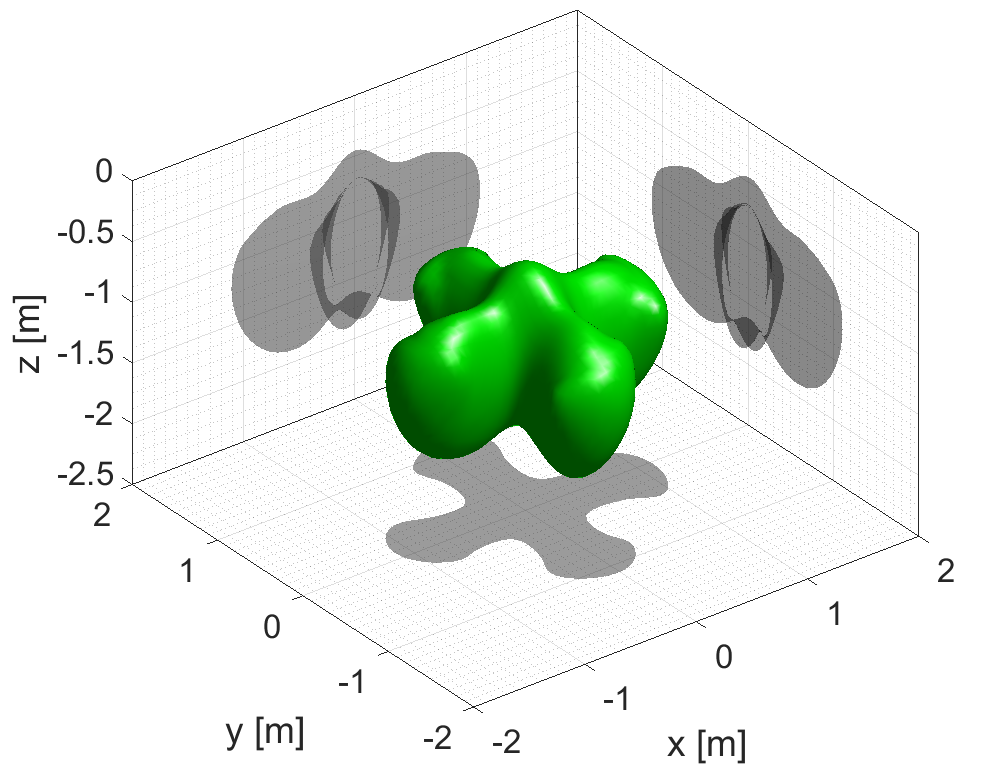}}\\
        \setcounter{subfigure}{2}\subfloat[]
        {\includegraphics[width=\dousize\linewidth] {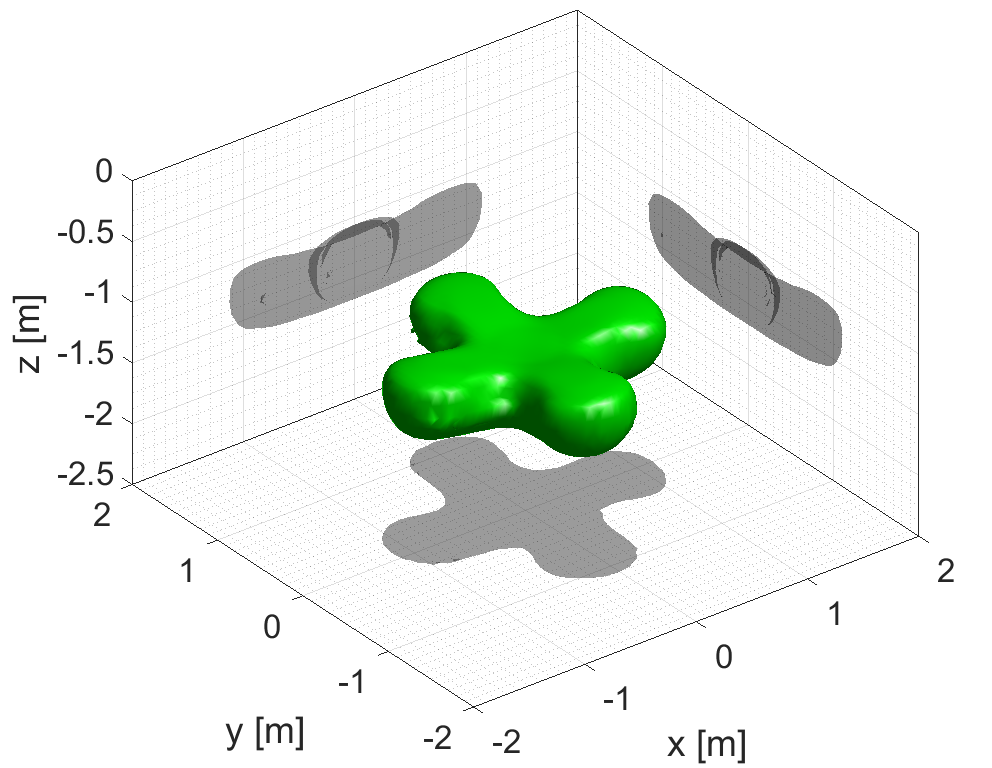}}\hspace*{\intsize cm}
        \setcounter{subfigure}{3}\subfloat[]
        {\includegraphics[width=\dousize\linewidth] {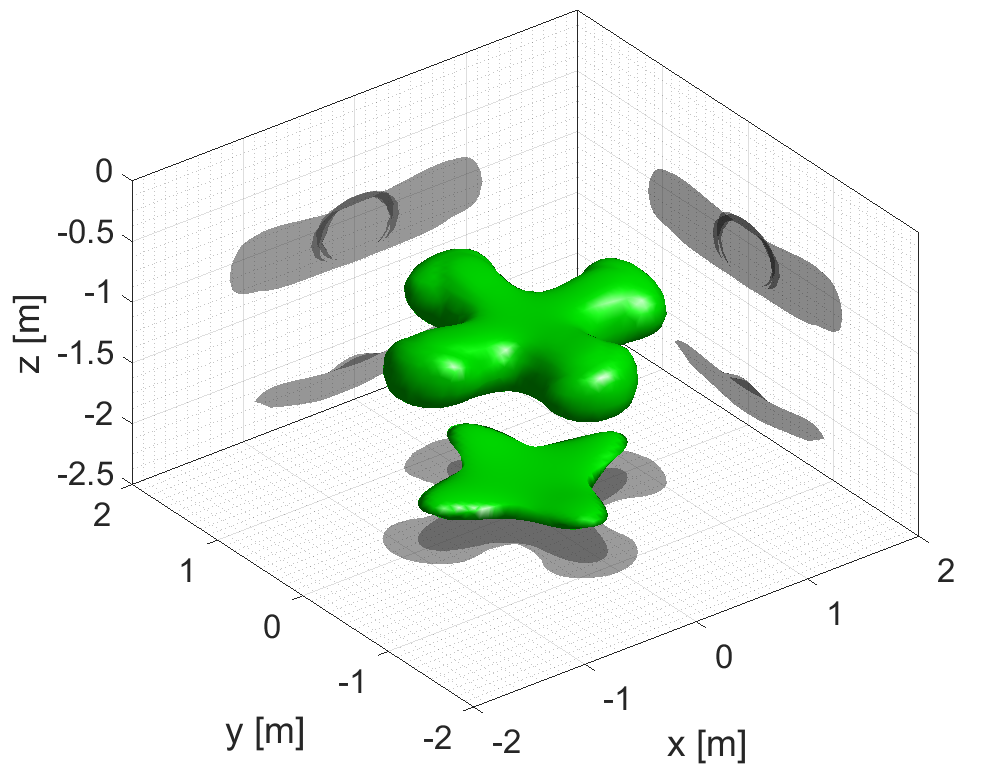}}\\
        \setcounter{subfigure}{4}\subfloat[]
        {\includegraphics[width=\dousize\linewidth] {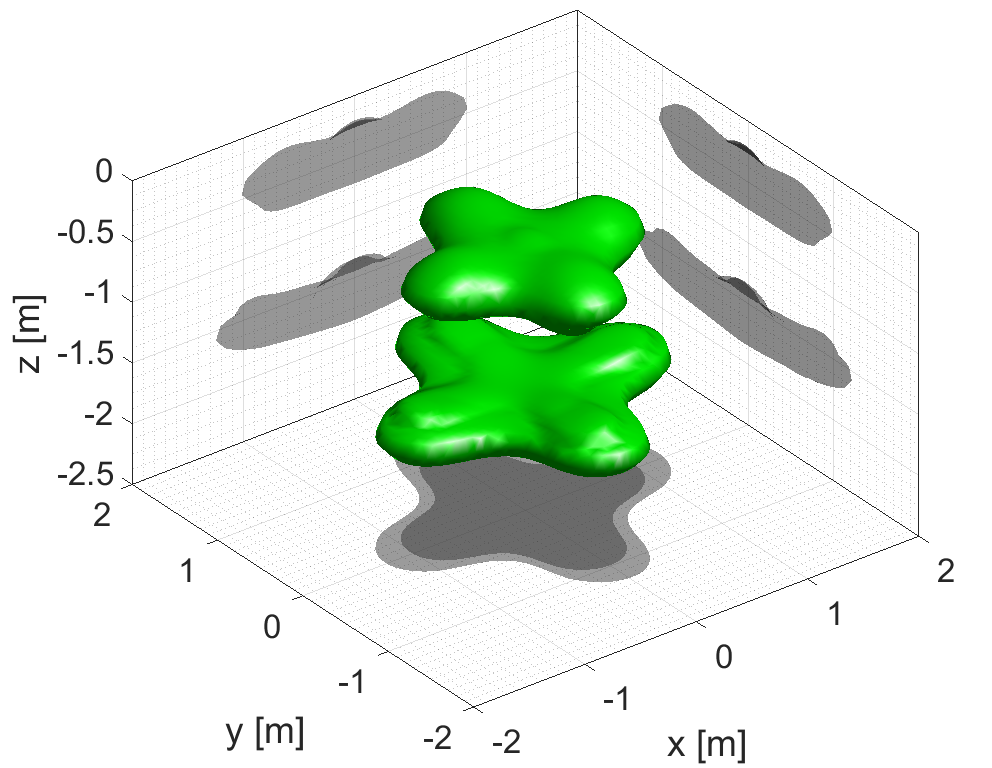}}\hspace*{\intsize cm}
        \setcounter{subfigure}{5}\subfloat[]
        {\includegraphics[width=\dousize\linewidth] {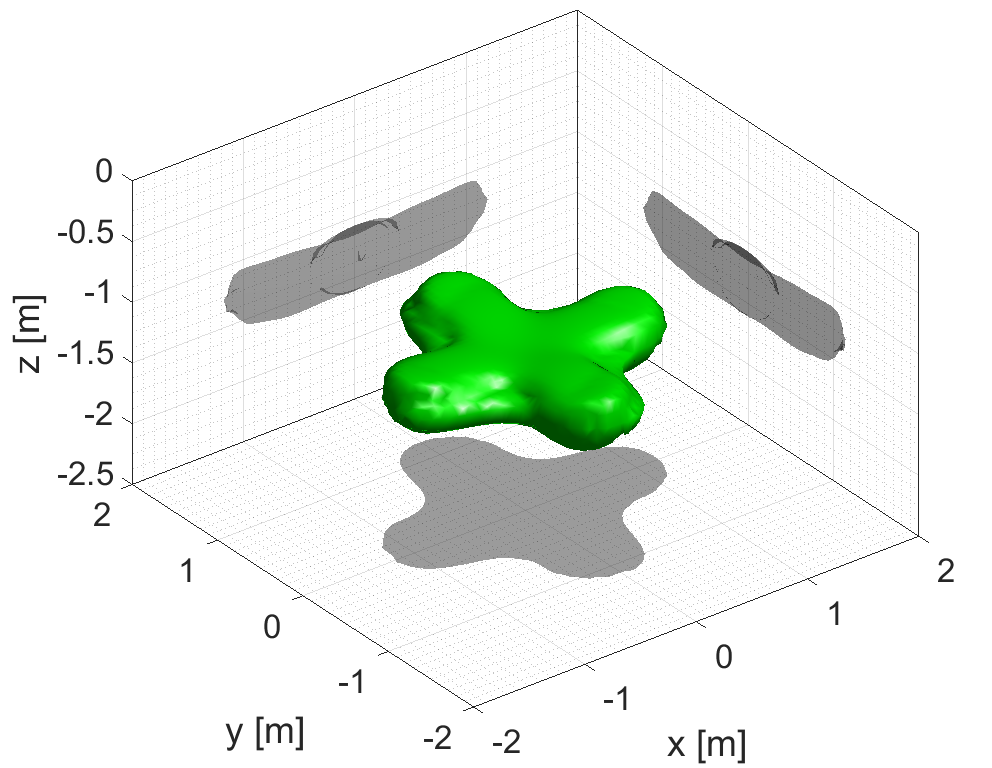}}
        \caption{Three-dimensional shape of the reconstructed results in the TW imaging experiment at 200 MHz. 5\% random white noise is added. (a) True objects. (b) Reconstructed contrast sources. (c) and (d) Reconstructed contrast permittivity and conductivity using exact background model. (e) and (f) Reconstructed contrast permittivity and conductivity using inexact background model ($0.75\epsilon_b$).}
        \label{fig:invwalldiesmshape}
    \end{figure}
    Let us first use the lossy cross object whose relative permittivity $\varepsilon_r=2$ and conductivity $\sigma=0.001$ S/m. The TW measurement data was disturbed by 5\% random white noise according to Eq.~\eqref{eq.noise}, and was then inverted with the incident fields and the scattering matrix calculated with the exact wall model. Fig.~\ref{fig:CVwalldie}(a) shows the reconstruction residual curve and the CV residual curve for recovering the contrast sources. The CV residual reaches the smallest at the iteration 90 where we obtain the optimal solution of the contrast sources. Fig.~\ref{fig:CVwalldie}(b) shows the data error curve and the state error curve for reconstructing the contrast. The iterative process converges after 10 iterations, however, there is a relatively large data error of 0.27 and a state error of 0.67 that cannot be minimized any more. This is due to the inexact estimation of the contrast sources which can be obviously seen from Fig.~\ref{fig:invwalldiesmshape}(a) and (b) in which the shape of the real object and the reconstructed contrast sources are shown. 

    \begin{figure}[!ht]
        \centering
        \setcounter{subfigure}{0}\subfloat[]
        {\includegraphics[width=\dousize\linewidth] {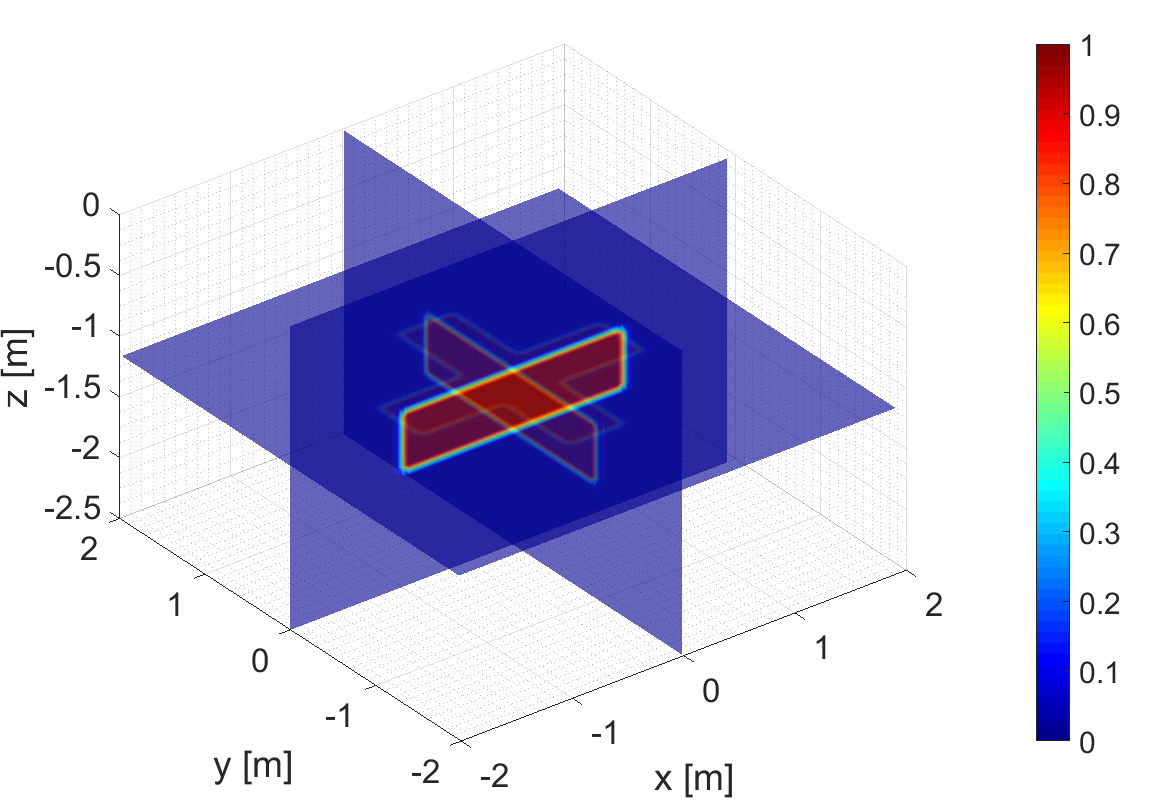}}\hspace*{\intsize cm}
        \setcounter{subfigure}{1}\subfloat[]
        {\includegraphics[width=\dousize\linewidth] {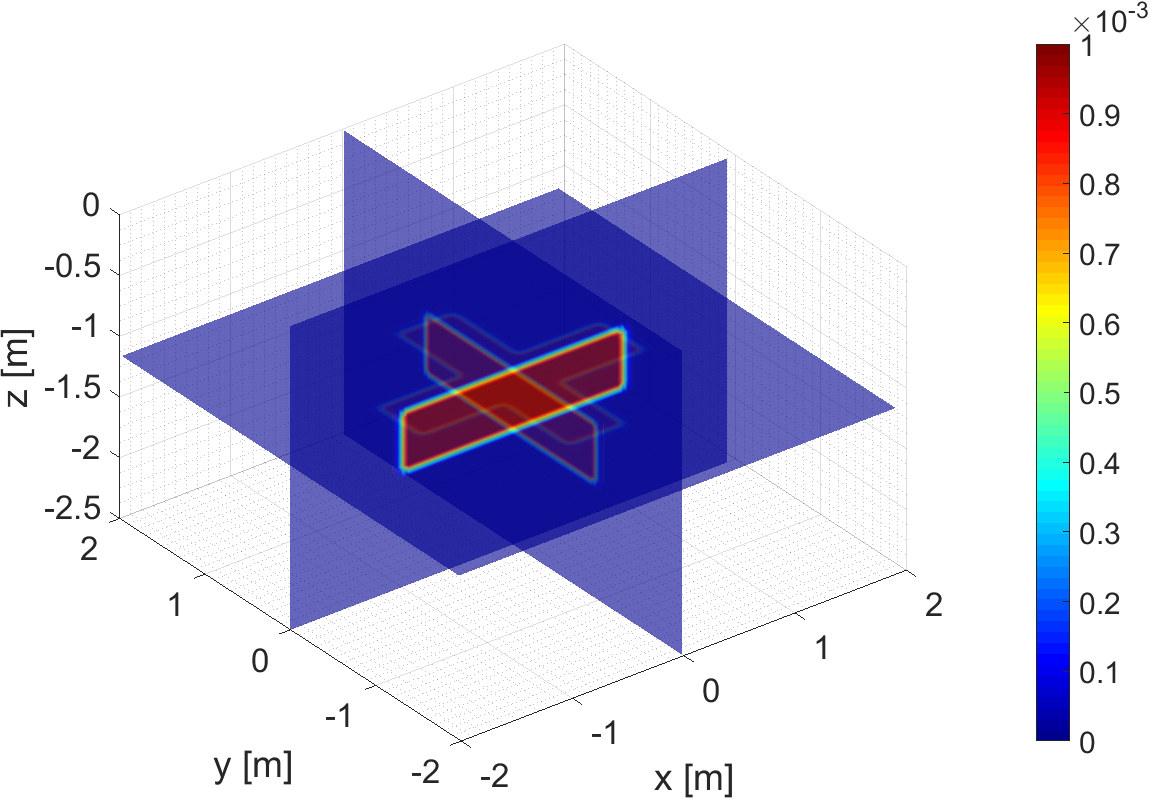}}\\
        \setcounter{subfigure}{2}\subfloat[]
        {\includegraphics[width=\dousize\linewidth] {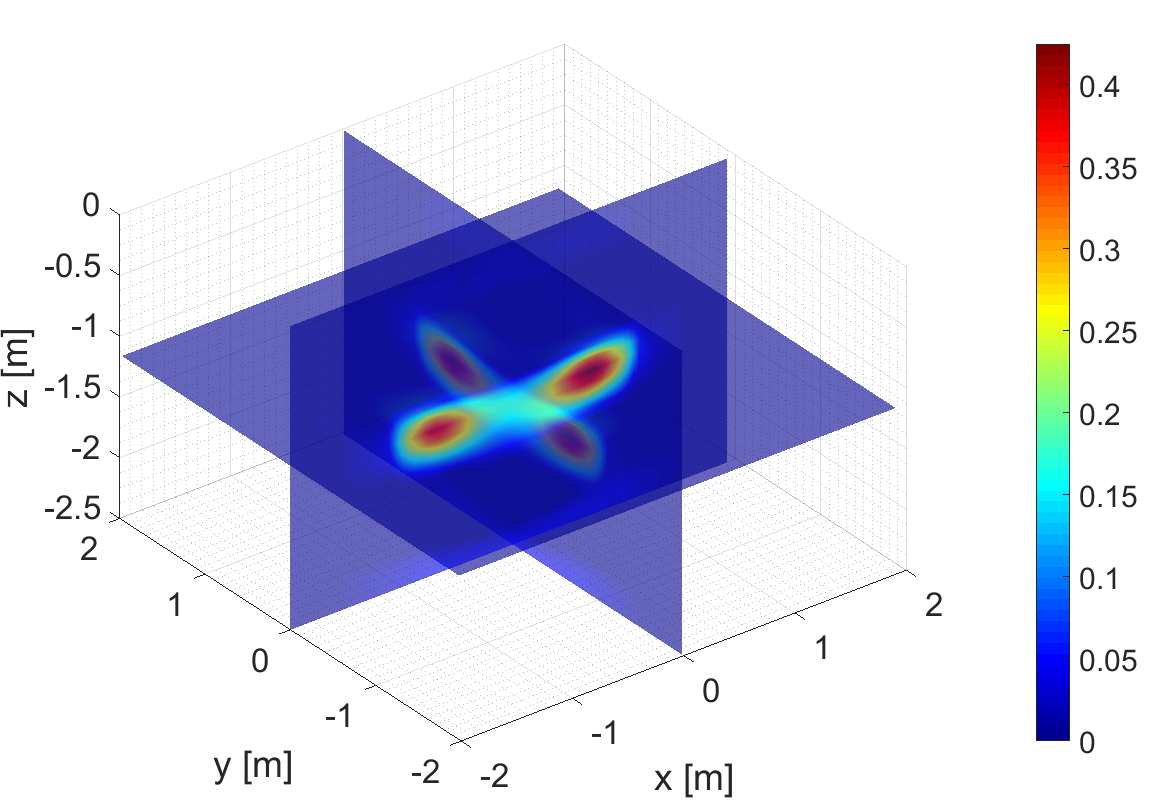}}\hspace*{\intsize cm}
        \setcounter{subfigure}{3}\subfloat[]
        {\includegraphics[width=\dousize\linewidth] {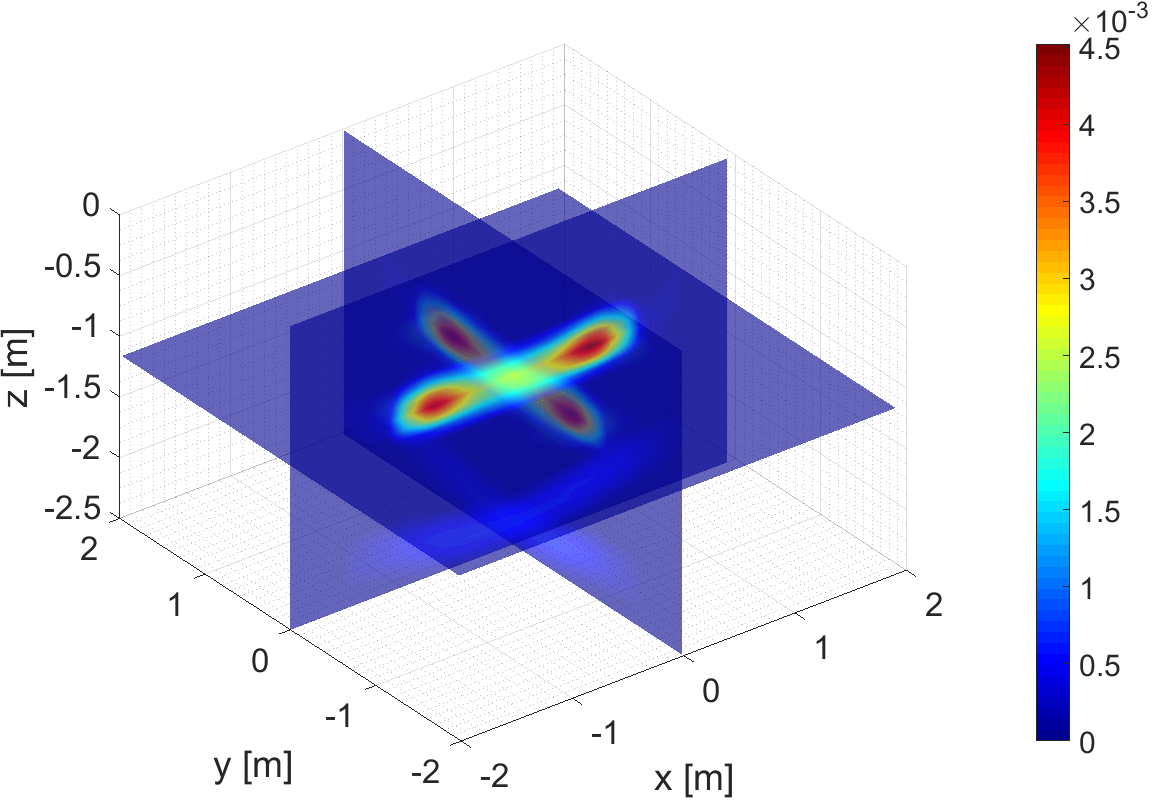}}\\
        \setcounter{subfigure}{4}\subfloat[]
        {\includegraphics[width=\dousize\linewidth] {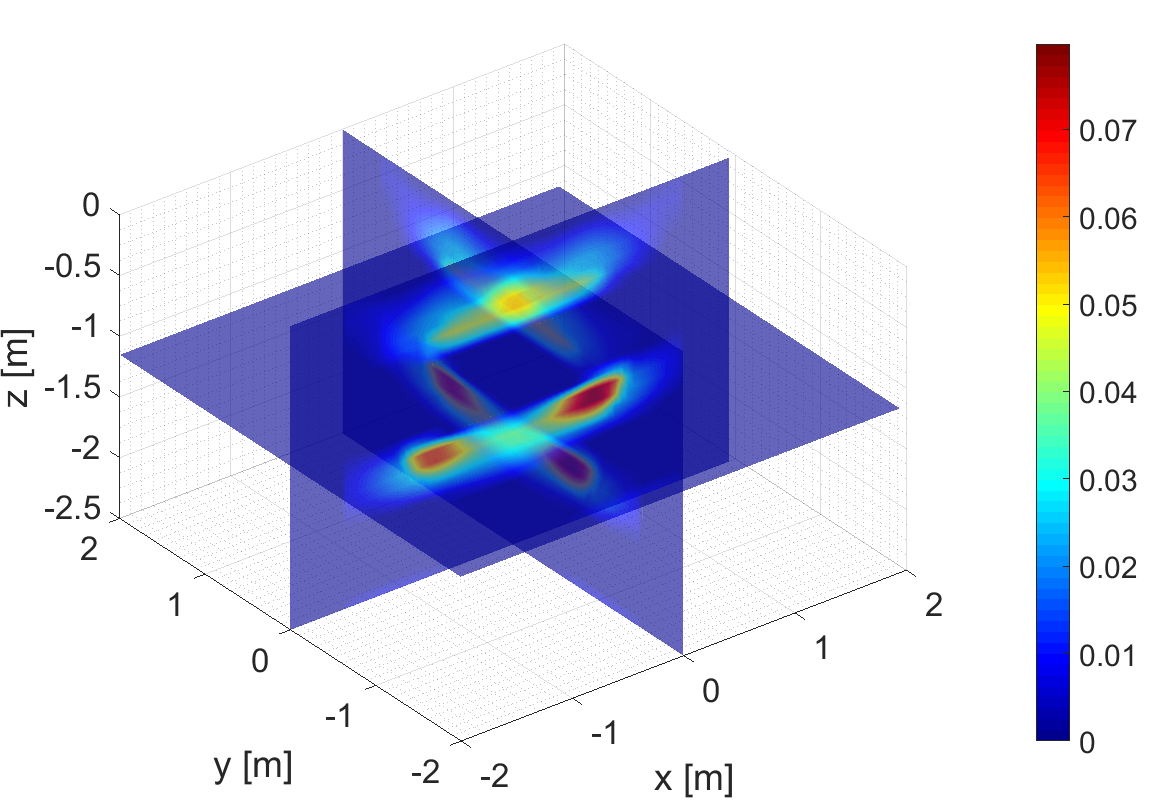}}\hspace*{\intsize cm}
        \setcounter{subfigure}{5}\subfloat[]
        {\includegraphics[width=\dousize\linewidth] {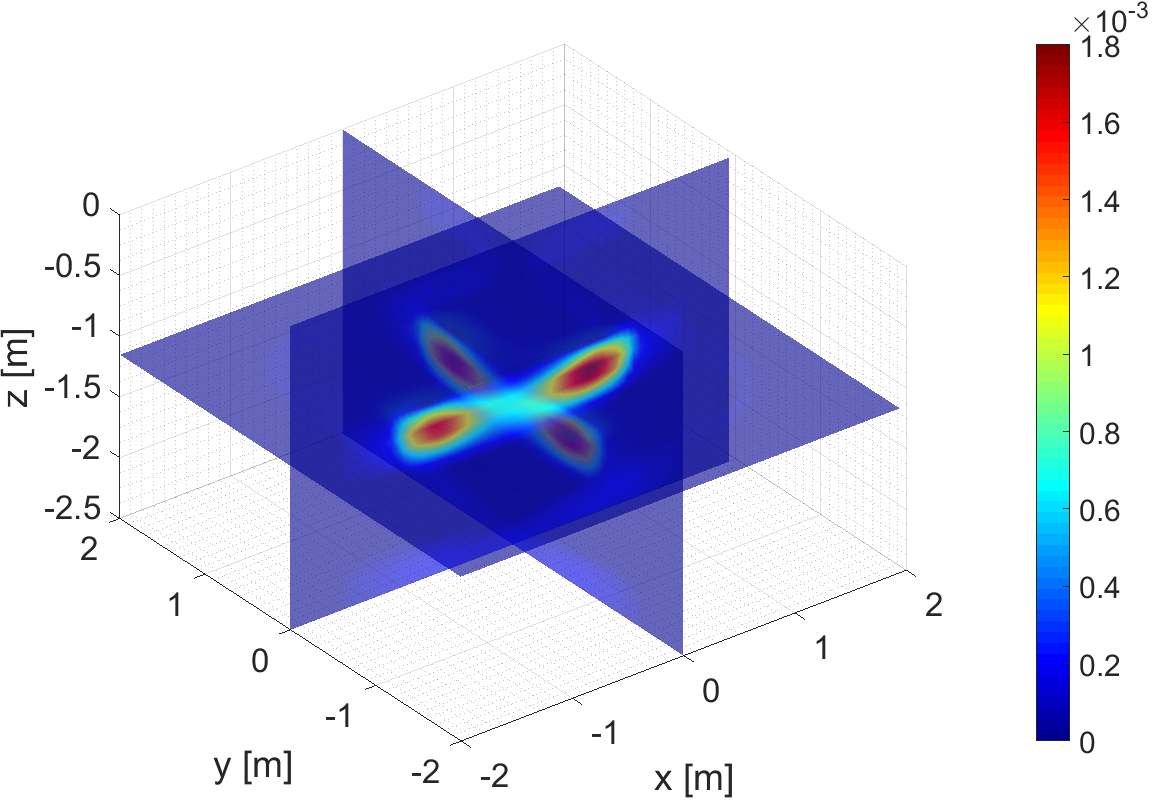}}
        \caption{Cross sections of the reconstructed dielectric parameters in the TW imaging experiment at 200 MHz. 5\% random white noise is added. The unit of the contrast conductivity is S/m. (a) True contrast permittivity. (b) True contrast conductivity. (c) and (d) Reconstructed contrast permittivity and conductivity using exact background model. (e) and (f) Reconstructed contrast permittivity and conductivity using inexact background model ($0.75\epsilon_b$).}
        \label{fig:invwalldie}
    \end{figure}
    As a matter of fact, due to the limited amount of independent measurement data, a good inversion can hardly be achieved with just the back-scattered fields. In our method, the contrast sources and the total fields are fixed while reconstructing the contrast, such that the iterative process can be prevented from converging to a local optimal solution which might be far away from the real solution. Fig.~\ref{fig:invwalldiesmshape}(c) and (d) give the shape of the reconstructed contrast permittivity and the reconstructed contrast conductivity, and the corresponding cross sections ($x=0$ m, $y=0$ m, and $z=-1.15$ m) are shown Fig.~\ref{fig:invwalldie}(c) and (d). For better comparison, the cross sections of the exact parameters are given as well in Fig.~\ref{fig:invwalldie}(a) and (b). We can see from Fig.~\ref{fig:invwalldiesmshape}(a), (c) and (d) that the shape of the cross is nicely reconstructed. The shown artefact in the reconstructed contrast conductivity is actually very weak compare to the reconstructed object, which can be seen from Fig.~\ref{fig:invwalldie}(d). If we average the reconstructed parameters in the cross region, then we have a coarse estimation of the contrast $\Delta\hat\varepsilon_r\approx 0.2$ and $\Delta\hat\sigma\approx 0.002$ S/m.  

    Assume that the dielectric parameters of the wall are underestimated by 25\%, the incident fields and the scattering matrix have to be recalculated correspondingly. However, since the object is surrounded by free space, the range constraints given by Eq.\eqref{eq.rangeconstr} keep the same. We do inversion to the same disturbed measurement data. Fig.~\ref{fig:CVwalldie}(c) shows the reconstruction residual curve and the CV residual curve for recovering the contrast sources, and Fig.~\ref{fig:CVwalldie}(d) shows the data error curve and the state error curve for reconstructing the contrast. The CV residual starts to increase at the iteration 79 where we obtain the optimal solution of the contrast sources. By comparison of Fig.~\ref{fig:CVwalldie}(b) and (d) we can see that the inexact wall model results in larger data error (0.45) and state error (1.65) compared to the one with the exact background model. The shape of the reconstructed contrast permittivity and the contrast conductivity with inexact wall model are shown in Fig.~\ref{fig:invwalldiesmshape}(e) and (f), respectively. And the corresponding cross sections are given in Fig.~\ref{fig:invwalldie}(e) and (f). An obvious ghost cross above the real object can be seen in the reconstructed contrast permittivity, while the reconstructed contrast conductivity is still good enough to identify the object. By comparison of Fig.~\ref{fig:invwalldie}(c) and (d) and (e) and (f), we see that after introducing the mismatch of the background, the reconstructed contrast conductivity stays at the same order of magnitude, while the reconstructed contrast conductivity is lower than the exact value 1 by around one order of magnitude.    

\subsubsection{Highly conductive object}

    \begin{figure}[!ht]
        \centering
        \subfloat[]
        {\includegraphics[width=0.45\linewidth]{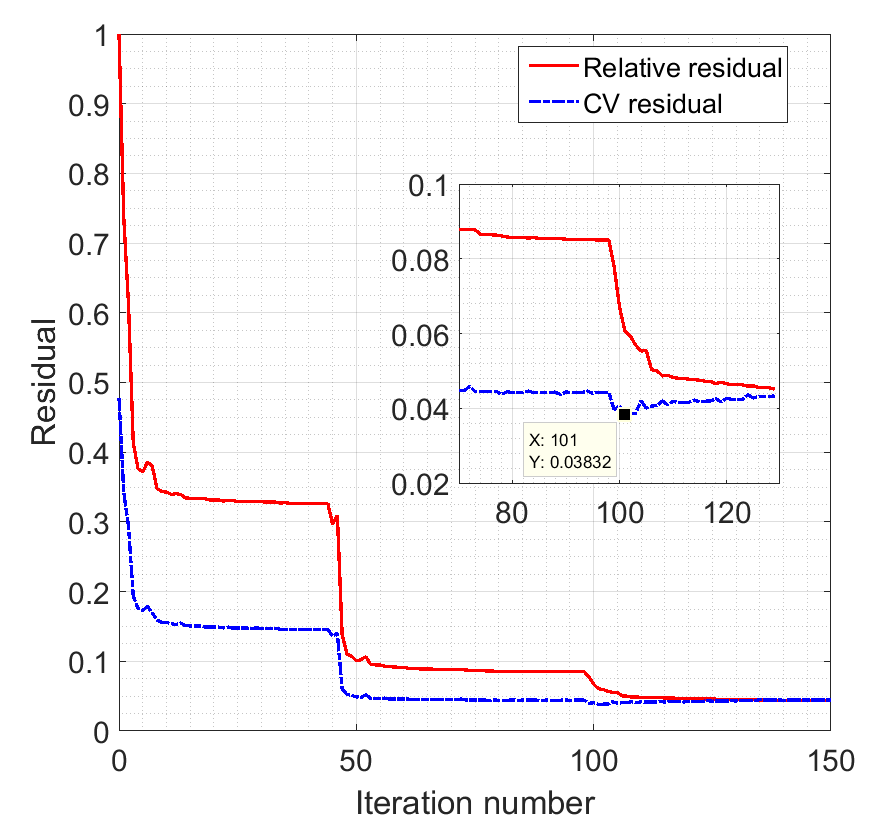}}\quad
        \subfloat[]
        {\includegraphics[width=0.45\linewidth] {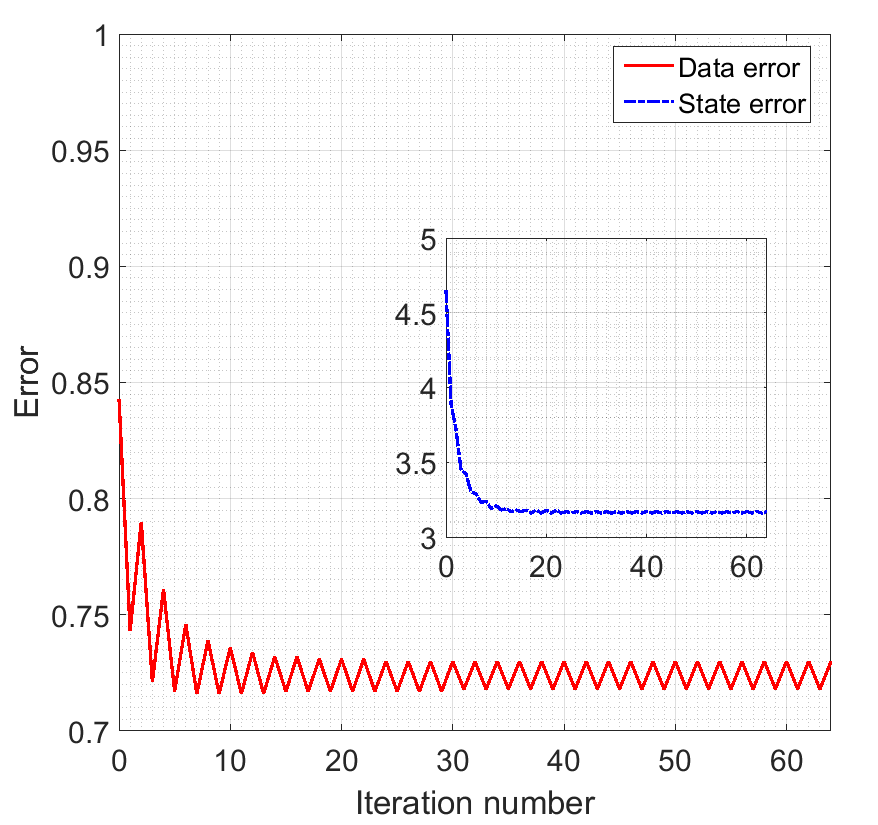}}\\
        \subfloat[]
        {\includegraphics[width=0.45\linewidth]{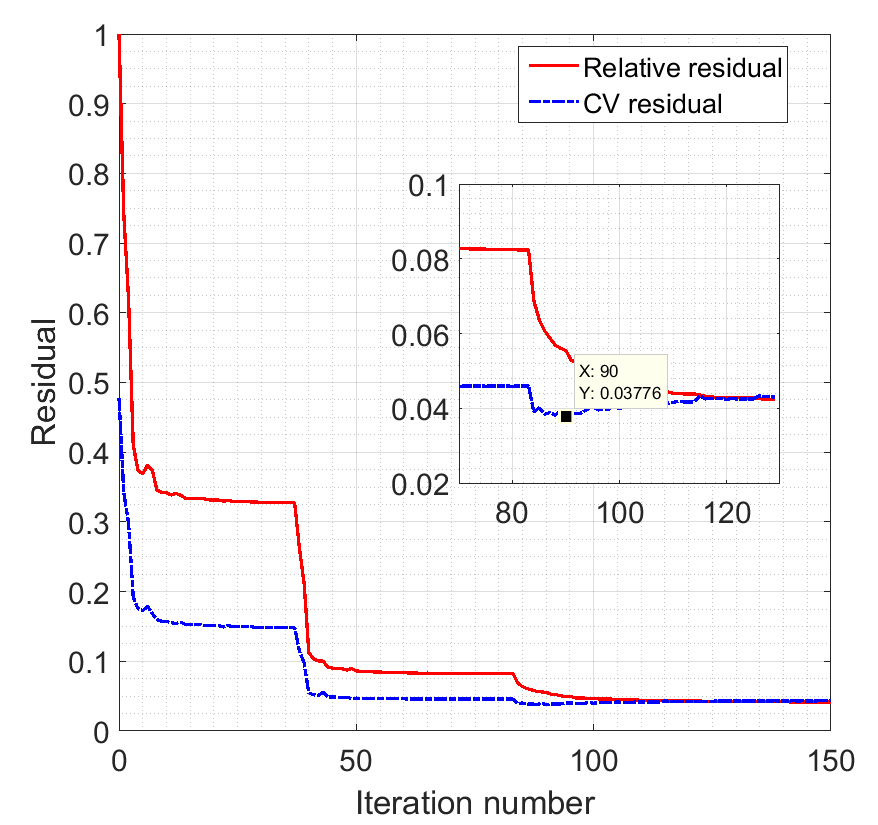}}\quad
        \subfloat[]
        {\includegraphics[width=0.45\linewidth] {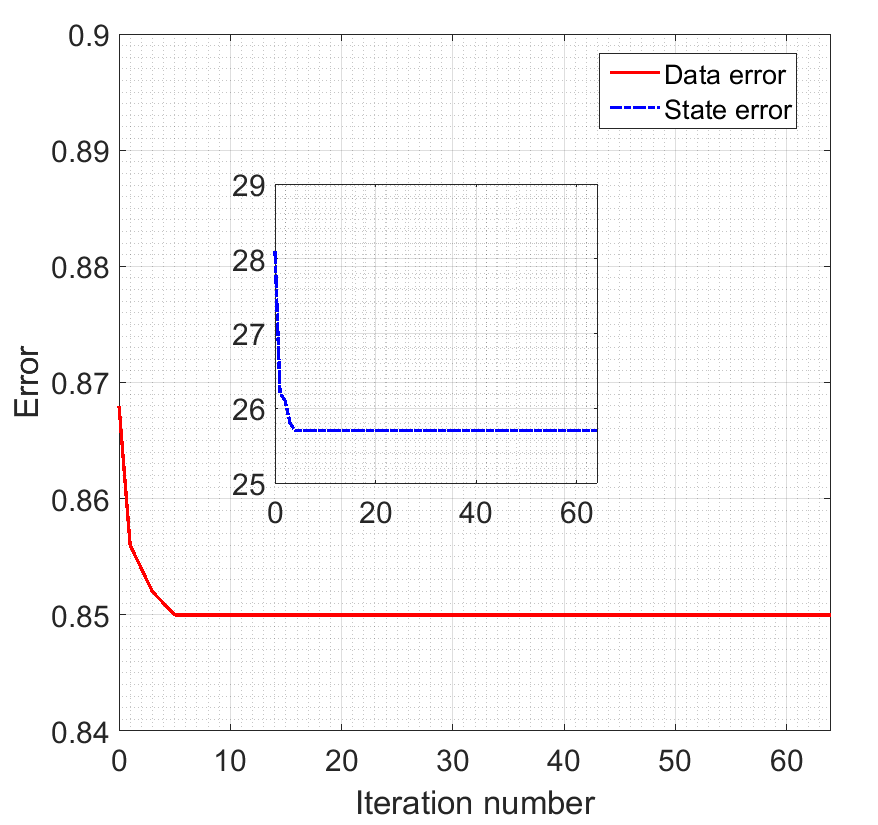}}
        \caption{Residual curves of the TW imaging experiment with highly conductive object. (a) and (c) Reconstruction residual and CV residual curves for estimating the contrast sources using exact background model and inexact background model (the thickness is $0.75$ m), respectively. (b) and (d) Data error and state error curves for reconstructing the contrast using exact background model and inexact background model (the thickness is $0.75$ m), respectively.}
        \label{fig:CVwallPEC}
    \end{figure}
    \begin{figure}[!ht]
        \centering
        \setcounter{subfigure}{0}\subfloat[]
        {\includegraphics[width=\dousize\linewidth] {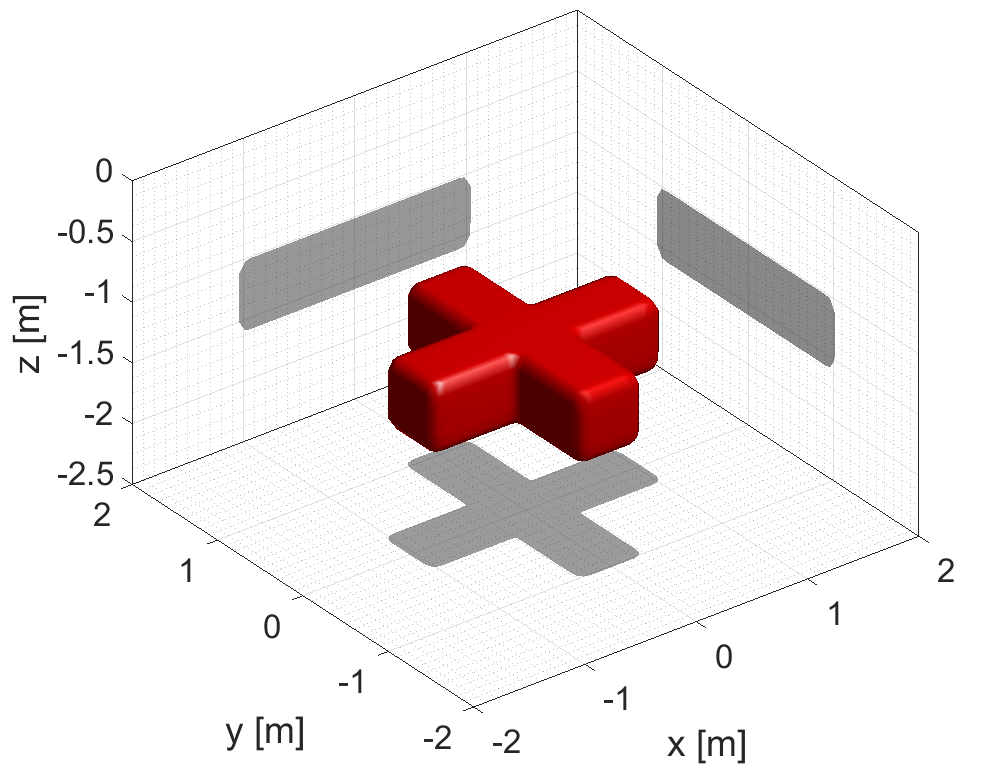}}\hspace*{\intsize cm}
        \setcounter{subfigure}{1}\subfloat[]
        {\includegraphics[width=\dousize\linewidth] {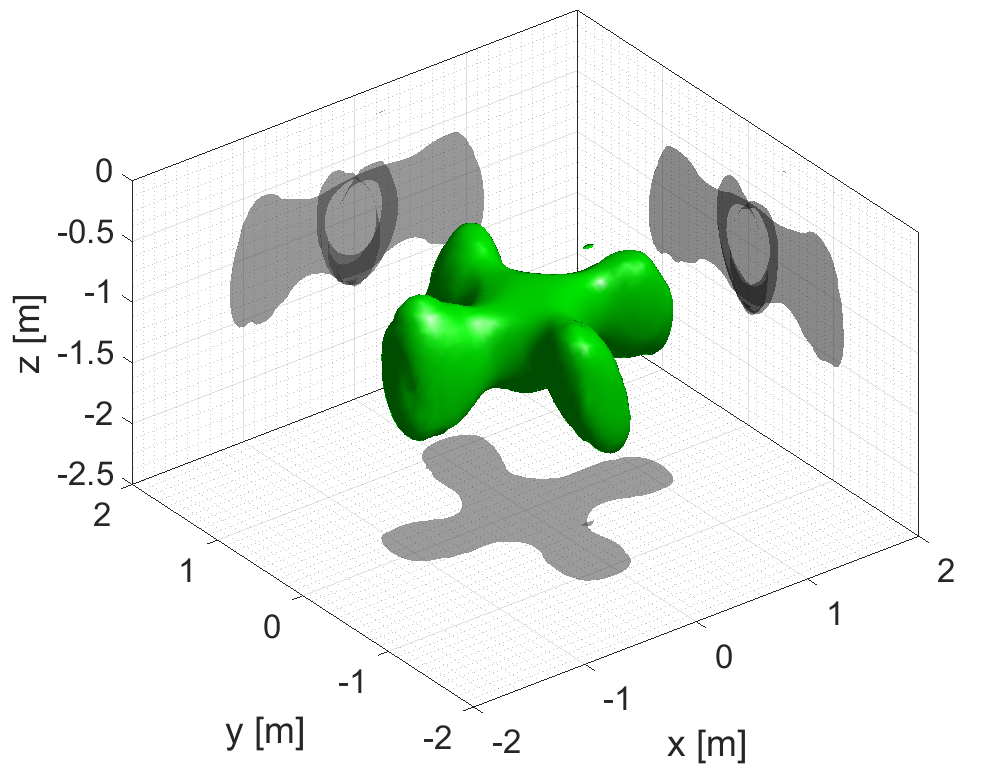}}\\
        \setcounter{subfigure}{2}\subfloat[]
        {\includegraphics[width=\dousize\linewidth] {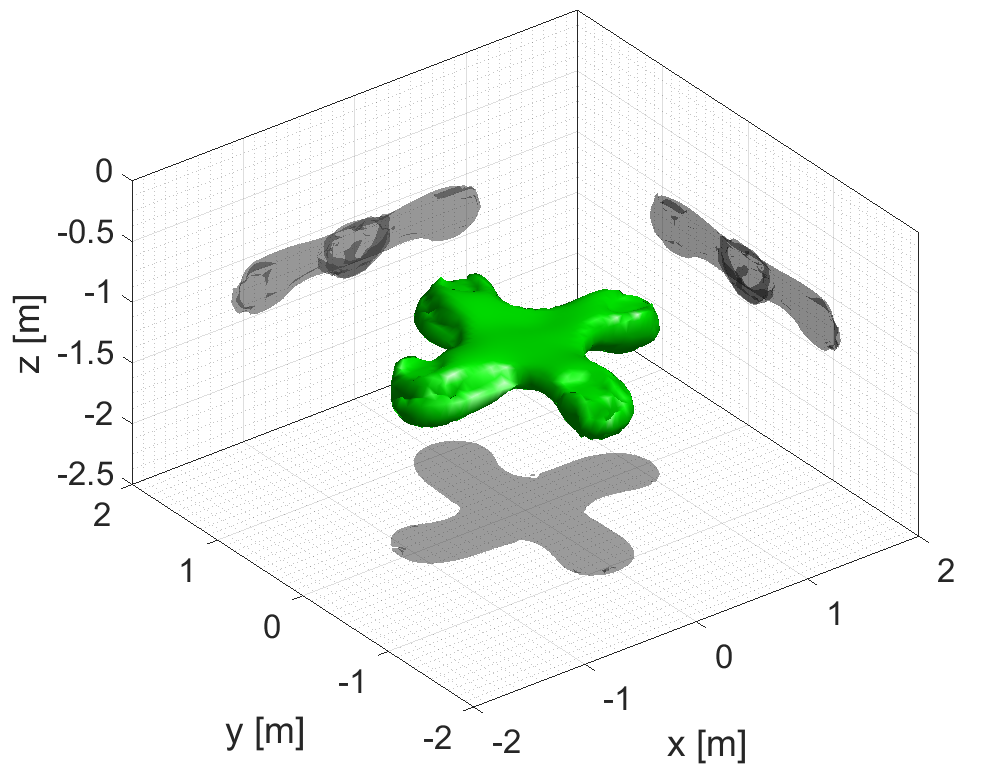}}\hspace*{\intsize cm}
        \setcounter{subfigure}{3}\subfloat[]
        {\includegraphics[width=\dousize\linewidth] {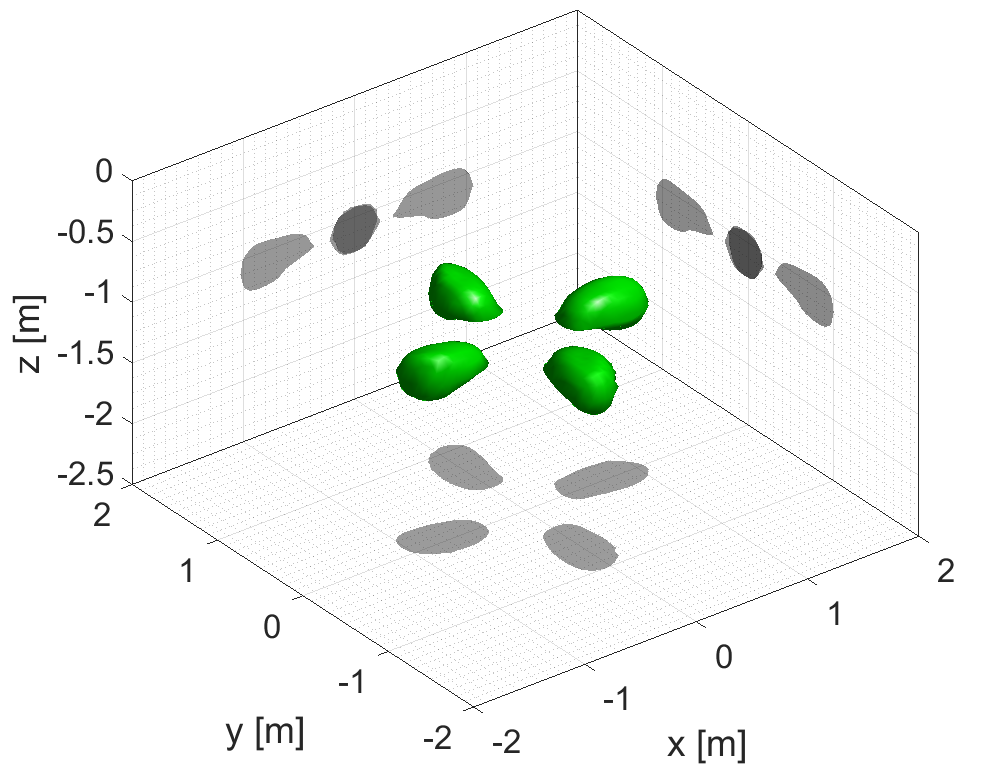}}\\
        \setcounter{subfigure}{4}\subfloat[]
        {\includegraphics[width=\dousize\linewidth] {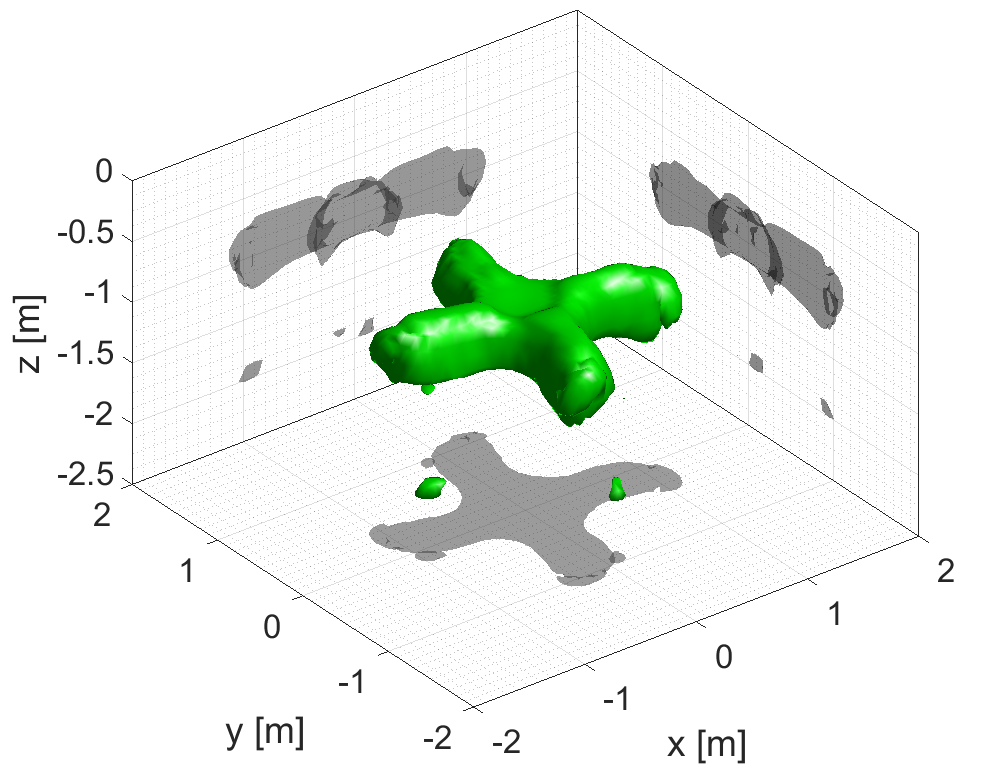}}\hspace*{\intsize cm}
        \setcounter{subfigure}{5}\subfloat[]
        {\includegraphics[width=\dousize\linewidth] {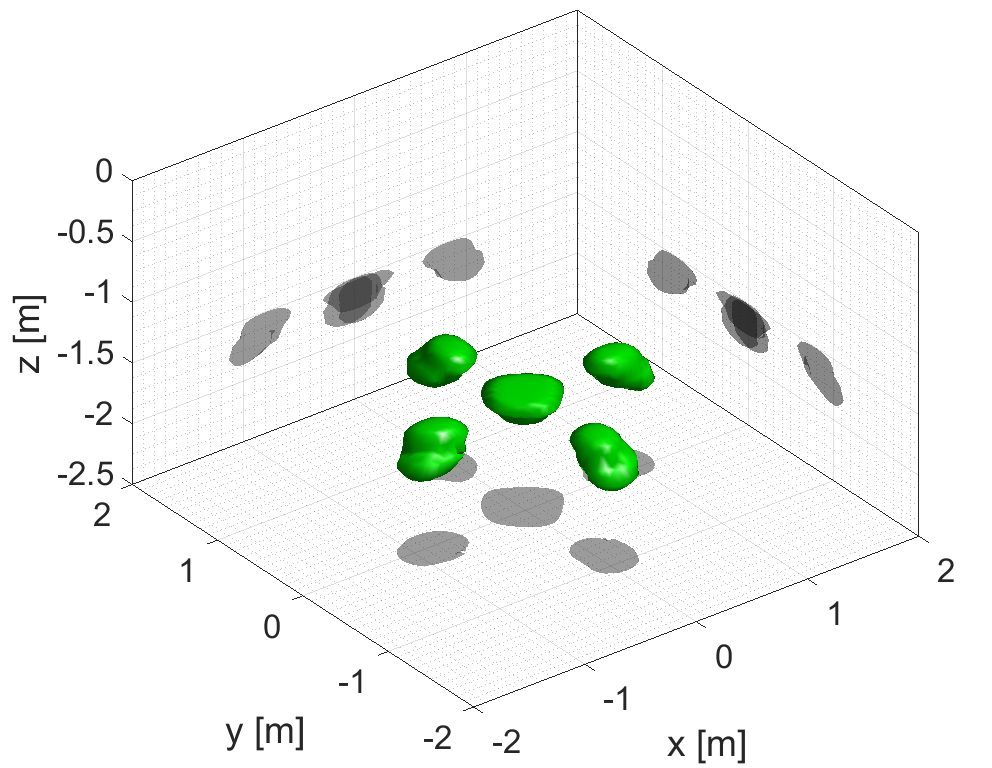}}
        \caption{Three-dimensional shape of the reconstructed results in the TW imaging experiment at 200 MHz. 5\% random white noise is added. (a) Real objects. (b) Reconstructed contrast sources. (c) and (d) Reconstructed contrast permittivity and conductivity using exact background model. (e) and (f) Reconstructed contrast permittivity and conductivity using inexact background model (the thickness is $0.75$ m).}
        \label{fig:invwallPECshape}
    \end{figure}
    \begin{figure}[!ht]
        \centering
        \setcounter{subfigure}{0}\subfloat[]
        {\includegraphics[width=\dousize\linewidth] {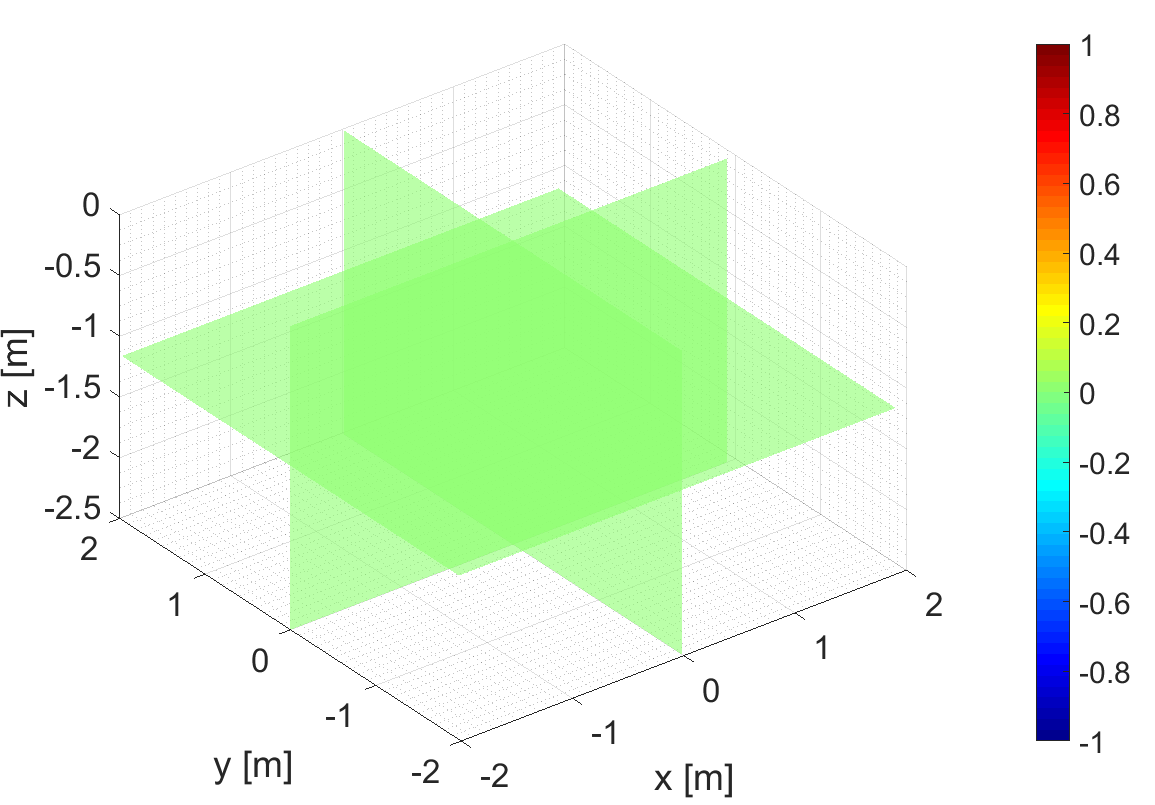}}\hspace*{\intsize cm}
        \setcounter{subfigure}{1}\subfloat[]
        {\includegraphics[width=\dousize\linewidth] {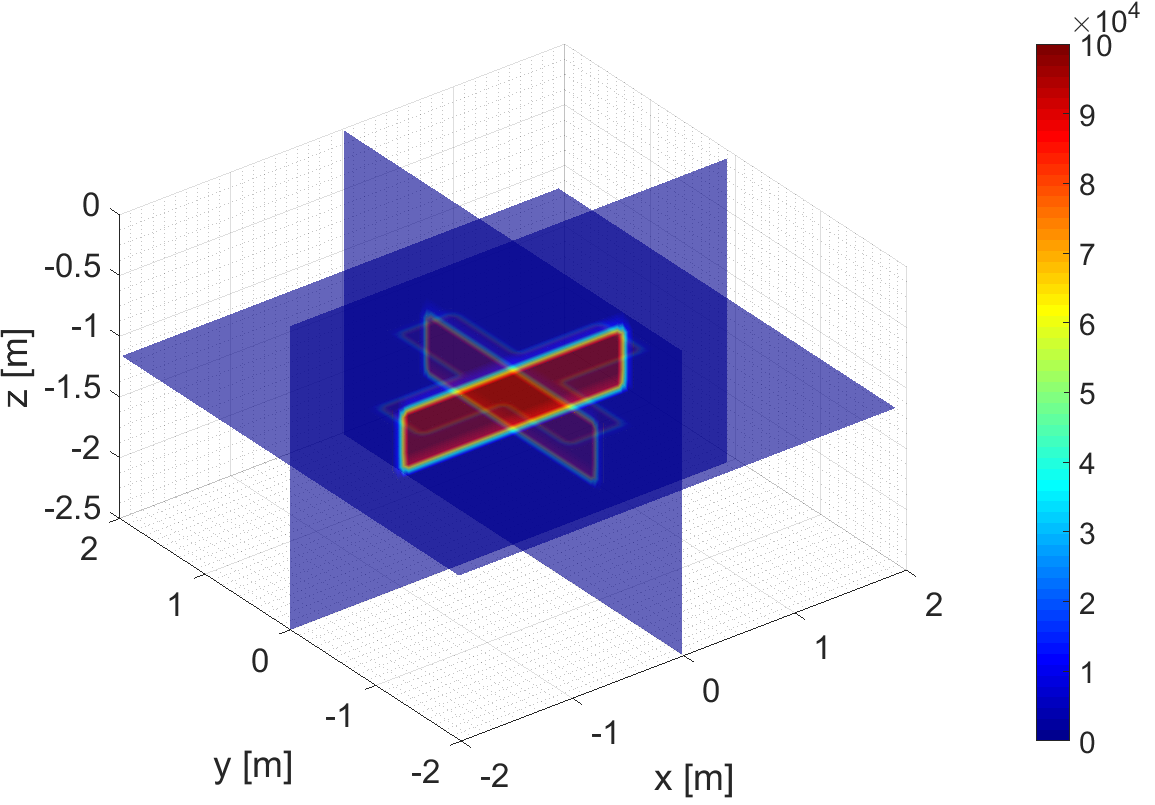}}\\
        \setcounter{subfigure}{2}\subfloat[]
        {\includegraphics[width=\dousize\linewidth] {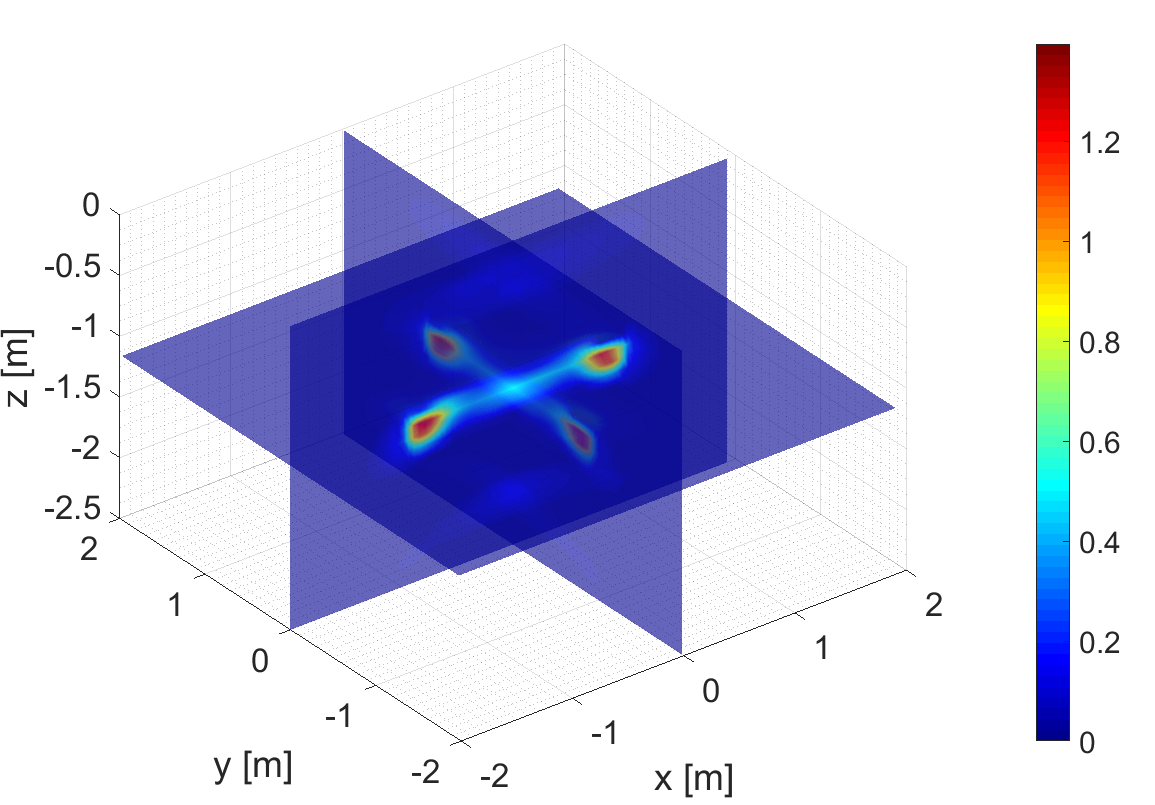}}\hspace*{\intsize cm}
        \setcounter{subfigure}{3}\subfloat[]
        {\includegraphics[width=\dousize\linewidth] {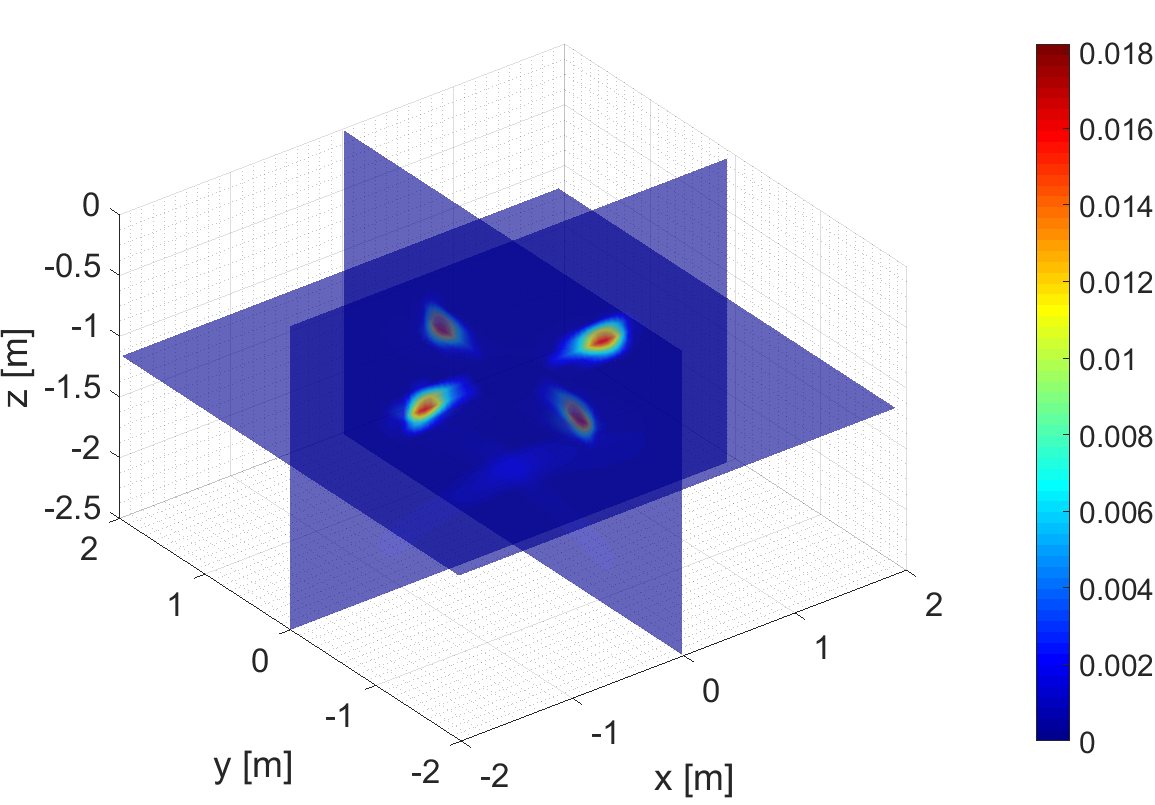}}\\
        \setcounter{subfigure}{4}\subfloat[]
        {\includegraphics[width=\dousize\linewidth] {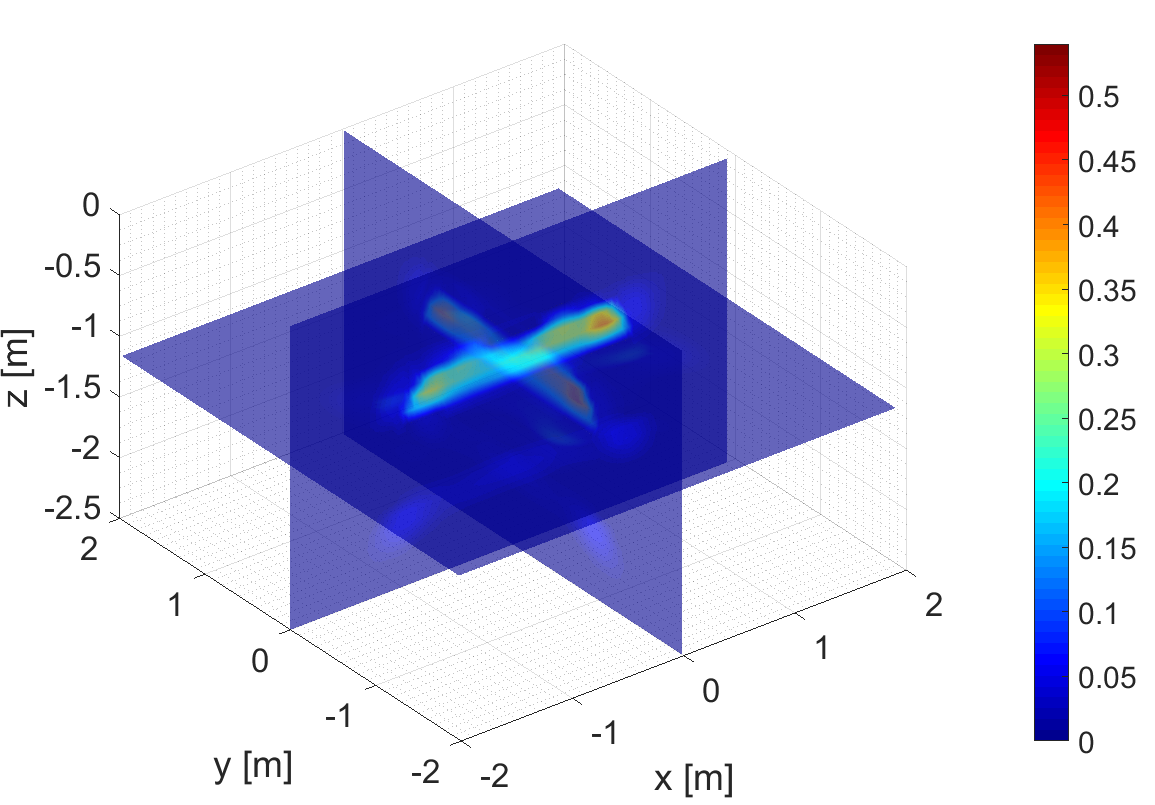}}\hspace*{\intsize cm}
        \setcounter{subfigure}{5}\subfloat[]
        {\includegraphics[width=\dousize\linewidth] {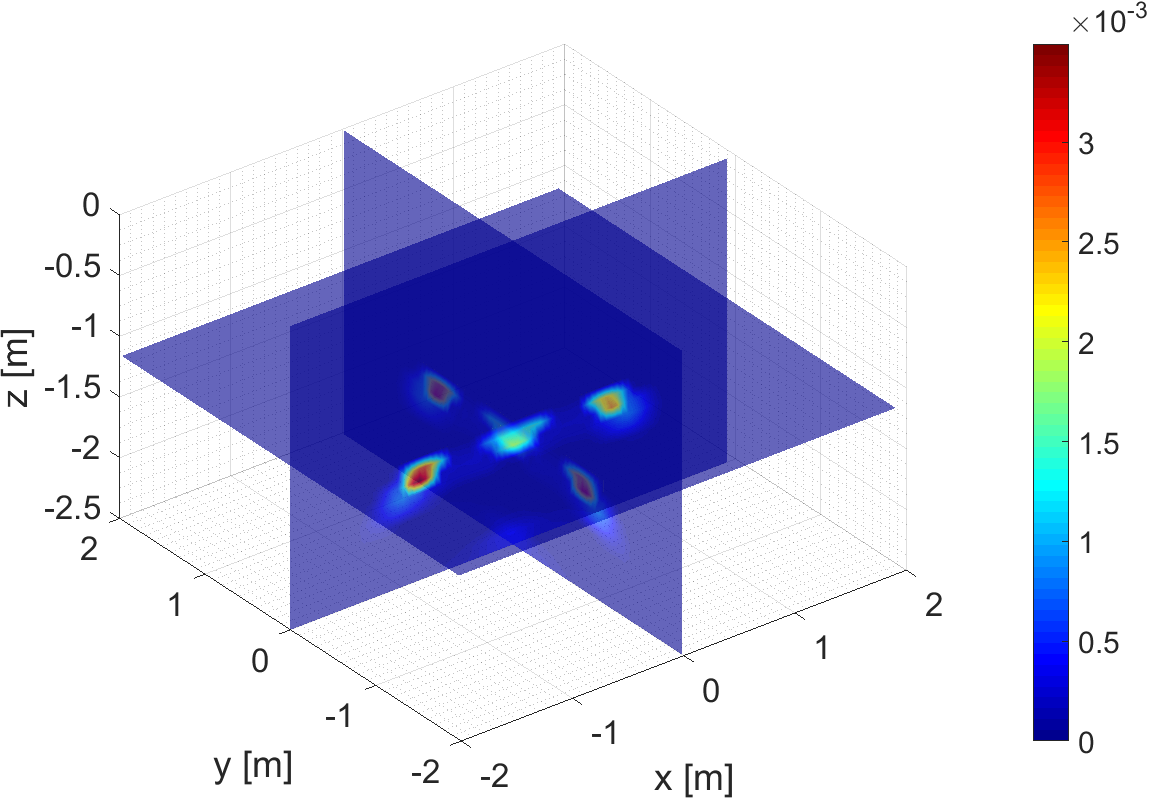}}
        \caption{Cross sections of the reconstructed dielectric parameters in the TW imaging experiment at 200 MHz. 5\% random white noise is added. The unit of the contrast conductivity is S/m. (a) True contrast permittivity. (b) True contrast conductivity. (c) and (d) Reconstructed contrast permittivity and conductivity using exact background model. (e) and (f) Reconstructed contrast permittivity and conductivity using inexact background model (the thickness is $0.75$ m).}
        \label{fig:invwallPEC}
    \end{figure}
    To study the performance of the proposed method for the inversion of highly conductive objects, let us now test a cross object of the same size but made of highly conductive material ($\varepsilon_r=1$, $\sigma=10^5$ S/m). Firstly, let us process the measurement data using the exact background model. Namely, we do the same thing as that of the previous lossy cross object, but only replacing the measurement data. Fig.~\ref{fig:CVwallPEC}(a,b) give the residual curves of recovering both the contrast sources and the contrast. We can see that the data error and the state error are larger than those of inverting the lossy cross object, indicating that the contrast sources are reconstructed with a larger error. Fig.~\ref{fig:invwallPECshape}(a) and (b) show the shape of real cross object and the shape of the inverted contrast sources, from which we can see that the contrast sources have a basic cross-like shape extending along the $z$-axis from $z_1=-0.5$ m to $z_2=-1.5$ m. As a matter of fact, for highly conductive objects, the contrast sources are supposed to distribute on the top surface of the object, and the EM fields in the interior are zero due to the serious attenuation. This nicely explains why the middle of the contrast sources is exactly the top surface of the real object, i.e., $(z_1+z_2)/2=-1$ m. The reconstructed contrast permittivity and contrast conductivity are shown in Fig.~\ref{fig:invwallPECshape}(c) and (d), and the corresponding cross sections are shown in Fig.~\ref{fig:invwallPEC}(c) and (d). We can see a ghost of the contrast permittivity is reconstructed with a maximum value of 1, and the reconstructed contrast conductivity is more focused on the top surface of the cross object due to the higher conductivity compared to the lossy cross object. Although the estimation of the dielectric parameters is not accurate in the inversion of highly conductive objects, we get the basic morphological information of the object in the inverted results, which is of more importance in real applications. 

    In the inversion with inexact wall model, we changed the thickness of the wall to 0.75 m while using the exact dielectric parameters. The residual curves are shown in Fig.~\ref{fig:CVwallPEC}(c) and (d). Obviously, the data error and the state error are larger than those of the inversion using exact wall model (see Fig.~\ref{fig:CVwallPEC}(a) and (b)). The corresponding inverted results are given in Fig.~\ref{fig:invwallPECshape}(e) and (f) and Fig.~\ref{fig:invwallPEC}(e) and (f), from which we can see that the contrast permittivity is lifted up by 0.25 m, and the contrast conductivity is lowered by 0.5 m. This is a very interesting phenomenon because the mismatch of the background model is reflected by the mismatch of the contrast permittivity and the contrast conductivity. 

\subsection{Numerical performance}

    To summarize this section, we remark that the linearized 3-D inversion method gives good inverted results for the inversion of lossy objects in GPR imaging and TW imaging. It is also able to provide the morphological information of highly conductive objects. The quality of the background estimation is critical for ensuring the accuracy of the inverted results. This method leaves a large data error and a large state error for solving the 3-D half-space inverse problems. However it helps to prevent the iterative process from converging to a totally false local optimal solution in the cases where only back-scattered fields are available.  

    In the numerical experiments, the codes for reconstructing the contrast sources and the contrast are written by MATLAB codes. We ran the codes on a desktop with one Intel(R) Core(TM) i5-3470 CPU @ 3.20 GHz, and we did not use parallel computing. The running time of each iteration is 3.1 s and 2.0 s respectively for the GPR case, and 3.5 s and 2.2 s respectively for the TW case. The codes of solving the total fields are written with ``C'' language and PETSc. We run the codes on a server with two Intel(R) Xeon(R) CPUs E5-2650 v2 @ 2.60GHz containing 16 cores totally. Parallel computing is used with 16 cores. The running time of solving the total field for each source is 48 s for the GPR case and 25 s for the TW case. In the GPR case, we ran 200 iterations for recovering the contrast sources and 20 iterations for recovering the contrast, the total running time is about 40 min. ($(200\times 3.1+20\times 2.0+36\times 48)/60$). In the TW case, we ran 150 iterations for recovering the contrast sources and 20 iterations for recovering the contrast, the total running time is about 27.4 min. ($(150\times 3.5+20\times 2.2+36\times 25)/60$). However, if we implement the MATLAB codes using parallel computing technique with 16 cores, the running time can be reduced by at least 8 times. In doing so, the total running time can be further reduced to 30 min and 16.5 min, respectively. While for the traditional iterative inversion methods, such as CSI and BIM, the update of the total fields in each iteration requires at least 57.6 min. ($(2 \times 36\times 48)/60$) and 30 min. ($(2 \times 36\times 25)/60$), respectively. The total running times are at least a multiple of the above times, where the multiple is determined by the required iteration number. This shows that the proposed method is far more efficient in comparison to the traditional iterative inversion methods. 

\section{Conclusion}\label{sec.Conclusion}
  
    In this paper, we have proposed to exploit the joint structure of the contrast sources in order to overcome the ill-posedness of the inverse scattering problem. The contrast sources are obtained by solving a linear sum-of-$\ell_1$-norm optimization problem, of which the objective is to obtain a regularized solution, instead of finding the sparsest solution. The inversion domain is discretized for the application of a finite-difference frequency-domain (FDFD) scheme to model the electromagnetic state equation in the scattering problem, resulting in a highly accurate scattering model, which can be applied to configurations with a versatile known background. In the cases where the scattering objects extending over a large region, the influence of the regularization constraint becomes less significant. However, this problem can be, to a very large extent, overcome by a multi-frequency version of the proposed method which is going to be discussed in another paper. With the estimated contrast sources, a linearized 3-D EM contrast source inversion method is further proposed. The proposed method is tested successfully in a ground penetrating radar (GPR) configuration and a through-the-wall (TW) configuration together with a 3-D FDFD solver. The inversion quality with both an exact and inexact background model are discussed and its sensitivity of the reconstruction in relation to the background medium estimation is shown. Results show that this method is not only efficient but also robust with respect to the reconstruction quality when the amount of the acquired measurement data is limited by the one-side probing configuration. In a following work, we will apply the proposed method to GPR and TW configurations using real measured data.

\bibliographystyle{IEEEtran}
\bibliography{mybib}

\end{document}